\documentclass[a4paper,fleqn,usenatbib,useAMS]{mpazh}
\usepackage{graphicx}	
\usepackage{amsmath}	
\usepackage{amssymb}	
\usepackage{multicol}        
\usepackage{bm}		

\usepackage{color}
\usepackage{pdflscape}
\usepackage[T1]{fontenc}
\usepackage{ae,aecompl}
\usepackage{newtxtext,newtxmath}
\usepackage{lscape}
\usepackage{mathtext}
\usepackage{float}
\usepackage{textcomp}
\usepackage{rotating}
\usepackage{dcolumn}
\usepackage{tabularx}
\usepackage{pdflscape}
\usepackage{url}
\usepackage{textcomp}

\renewcommand{\theenumi}{\roman{enumi}}%

\newcommand{\obs}{\textit{ObsID}}
\newcommand{\xmm}{\textit{XMM-Newton}}
\newcommand{\chandra}{\textit{Chandra}}
\newcommand{\siK}{Si\,XIII K$_{\alpha}$}
\newcommand{\siLy}{Si\,XIV Ly$_{\alpha}$}
\newcommand{\feI}{Fe\,I K$_{\alpha}$}
\newcommand{\feK}{Fe\,XXV K$_{\alpha}$}

\newcommand{\feLy}{Fe\,XXVI Ly$_{\alpha}$}
\newcommand{\niK}{Ni\,XXVII\,K$_{\alpha}$}
\newcommand{\niLy}{Ni\,XXVIII Ly$_{\alpha}$}

\title[SS~433 X-ray Jets Diagnostics]{Diagnostics of Parameters for the X-ray Jets of SS~433 from High-Resolution Chandra Spectroscopy}
\author[Medvedev~et~al.]{Pavel~S.~Medvedev$^{1}$\thanks{E-mail: \href{mailto:tomedvedev@iki.rssi.ru}{tomedvedev@iki.rssi.ru}}, Ildar~I.~Khabibullin$^{2,1}$\thanks{E-mail: \href{mailto:khabibullin@iki.rssi.ru}{khabibullin@iki.rssi.ru}}, Sergey~Yu.~Sazonov$^{1}$
 \\
$^{1}$ Space Research Institute of the Russian Academy of Sciences (IKI), 84/32 Profsoyuznaya Str, Moscow, Russia, 117997 \\
$^{2}$ Max Planck Institute for Astrophysics, Karl-Schwarzschild-Strasse 1, 85741 Garching, Germany}
\date{Accepted XXX. Received YYY; in original form ZZZ}

\pubyear{2019}

\begin{document}

\label{firstpage}
\pagerange{\pageref{firstpage}--\pageref{lastpage}}
\maketitle

\begin{abstract}
The X-ray spectrum of the Galactic microquasar SS~433 contains a rich set of emission lines
of highly ionized atoms of heavy elements whose significant Doppler shift leaves no doubt that they are produced in collimated relativistic jets of outflowing material. We have performed a systematic analysis of the high-resolution X-ray spectra obtained by the Chandra observatory to determine the parameters of the jets within the multitemperature model of their emission that self-consistently predicts the source's line and continuum spectrum. The spectrum of SS~433 at energies below 3 keV is shown to be statistically satisfactorily described by the jet emission model, while the introduction of an additional hard component is required above 3 keV. We summarize the jet parameters (bulk velocity, opening angle, kinetic luminosity, base temperature, and relative elemental abundances) derived by fitting the data below 3 keV and describe the revealed degeneracies and systematic effects due to the presence of an additional component. Using the derived parameters, we show that the hard component is compatible with the emission from the hot (up to 40 keV) extension of the visible part of the jets moderately absorbed ($N_H \sim 2 \times 10^{23}$ cm$^{-2}$) in the cold-wind material. The combined X-ray emission model constructed in this way allows the broadband spectrum of SS~433 to be described self-consistently.
\end{abstract}
\section{Introduction}
\label{s:intro}
One of the most important predictions of the theory of disk accretion onto compact objects \citep{SS1973} is the launching of intense gas outflows in the form of massive winds and relativistic jets in the case where the mass transfer rate is close to or exceeds the critical level. Such outflows facilitate the self-regulation of the accretion process in this regime, so that the rate of mass inflow into the innermost disk regions, where the main energy release occurs, remains close to the critical one, with much of the released energy being converted into the kinetic energy of the wind and jets. On the whole, recent numerical simulations \citep{Ohsuga2011, Fender2014, Sadowski2014, Jiang2017} confirm this picture. However, the validity of various quantitative predictions, such as the total radiative accretion efficiency, the degree of collimation of the emission, and its spectrum, and the fraction of the energy carried away by the relativistic jets, can be clarified only through observations of the sources in which the regime of supercritical accretion is believed to be realized.

The class of such sources is fairly wide, for example, extremely bright states of X-ray binaries in the Galaxy, ultraluminous and supersoft ultraluminous X-ray sources in nearby galaxies\citep{Feng2011, Urquhart2016, Liu2015}, events of tidal disruption of stars by supermassive black holes at the centers of galaxies \citep[see][ and references therein]{Middleton2018}, and, possibly, growing massive black holes in the early Universe. In view of their extreme brightness, such sources can exert a noticeable influence on their environment, be they the molecular clouds at the centers of galaxies, the circum- or intergalactic gas, or the galactic medium of the first galaxies  \citep[see, e.g., ][]{Sazonov2018}. A sufficiently accurate description of this influence also requires knowing the quantitative characteristics of their emission, winds, and relativistic jets.

Ultraluminous X-ray sources (ULXs), i.e., compact extragalactic sources with X-ray luminosities $10^{39}$--$10^{41}$ erg s$^{-1}$, are apparently an especially important class of supercritical accretors \citep[or a review,see ][]{Feng2011}. Such sources dominate in the total X-ray luminosity of star-forming galaxies in the local Universe \citep{Mineo2012, Sazonov2017}  and may have played a prominent role in heating the early Universe before the reionization epoch \citep{Sazonov2017, Madau2017}.

The prototype of ultraluminous X-ray sources in our Galaxy is believed to be the unique microquasar SS~433 \citep{Fabrika2001, Begelman2006, Poutanen2007, Fabrika2015}, which has stably launched precessing relativistic jets of material (close in composition to the ordinary stellar gas) at least over more than 40 years of direct observations \citep[for a review, see ][]{Fabrika2004}. The mass transfer rate by the donor star is estimated to be 10$^{-4}$ M$_{\odot}$ yr$^{-1}$ based on the observed flux in the H$\alpha$ line \citep{Shklovsky1981, Heuvel1981}, which exceeds the critical level for any reasonable mass of the compact object (estimated to be in the range from 1 to 15~$M_{\odot}$ \citealt{Fabrika1990, Hillwig2004, Blundell2008, Kubota2010, Goranskij2011, Cherepashchuk2018}) at least by several hundred times, leaving the question about its nature open.

Apart from intense ultraviolet radiation, whose luminosity is estimated to be  $10^{40}$  erg s$^{-1}$  \citep{Dolan1997, Waisberg2018},  the relativistic jets
are the main observed manifestation of the operation of a supercritical ``central engine'', because its radiation (with a luminosity close to the Eddington one) must be efficiently screened from us by a thick
accretion disk and its wind. An observer located in a direction close to the disk axis could possibly see this radiation and SS~433 would actually resemble a ULX for this observer. However, the constraints obtained
by searching for the reflected collimated emission from SS~433 by the atomic and molecular gas in the Galactic plane do not confirm this picture \citep{KS2016} and suggest that SS~433 more likely resembles supersoft ULXs, with almost all of their radiation being emitted at energies below 1 keV \citep{Urquhart2016, Liu2015}.

A significant ($z \sim 0.1$) and periodically changing (due to the jet precession with an amplitude of 21\textdegree\ and a period of 162 days) Doppler shift allows one
to efficiently identify the jet emission lines in the source's optical spectrum and to accurately determine their parameters, primarily the velocity and the kinetic luminosity. They show a remarkable stability
over the entire period of available observations  \citep{Cherepashchuk2018}, with the probable exception of the jet ``switch-off'' periods observed at the times of enhanced radio and optical flaring activity
of the source \citep{Kotani2006}. The fact that the optical emission region corresponds to $\approx 0.5$--1 day of jet gas flight, i.e., lies at distances $\approx 3$--$6\times 10^{14}$ cm from the central source \citep{Fabrika1987, Panferov1993, Waisberg2018} that significantly exceed the system's characteristic sizes $\sim  10^{12}$ cm \citep[see, e.g.,][]{Hillwig2004}, should be taken into account.

In contrast to the optical emission, the X-ray emission from the system is determined largely by the relativistic jets, as evidenced by the shifted (like the
optical lines) intense emission lines of highly ionized atoms of heavy elements (neon, silicon, sulfur, iron, and nickel) observed in the X-ray spectrum \citep{Abell1979, Brinkmann1996, Kotani1996, Marshall2002, Lopez2006, Marshall2013a}. The X-ray emission from the jets is well described by the standard model of a nearly ballistic, moderately relativistic gas flow, which becomes visible to a distant observer when its characteristic temperature is typically $T_0\sim 30$ keV \citep{KMS2016, Medvedev2018}. As one recedes from the base, the material cools down through adiabatic expansion and energy losses by radiation until the temperature reaches  $T \sim 0.1$ keV. Thereafter, a thermal instability apparently develops in the gas and causes fragmentation of the flow  \citep{Brinkmann1996}. An analysis of the orbital-precessional variability of the X-ray emission leads to the conclusion about close proximity of the radiative cooling region of the X-ray jets to the central sources, at distances $\sim 10^{12}$  cm \citep{Filippova2006, Marshall2013a}.  Therefore, determining the quantitative characteristics of the X-ray jets is also important for understanding the physical conditions in the system's more compact regions, in which the jets are formed, collimated, and accelerated ($<10^9$ cm).

The X-ray spectrum of the jets is a combination of continuum thermal bremsstrahlung radiation  coming from the hottest parts and a set of emission lines of the
above-mentioned elements forming predominantly in regions with temperatures that provide a maximum plasma emissivity in a given line \citep{KS2012}. 
This standard multitemperature thermal emission model  \citep{Brinkmann1988, Kotani1996, KMS2016}  turned out to be capable of satisfactorily describing the soft part (up to 3 keV) of the X-ray spectrum for SS 433.
However, it has soon become clear that the observed spectrum has an appreciable emission excess at higher energies. An additional component in the X-ray spectra of SS~433, which is not reproduced in the standard emission model, was identified by \cite{Brinkmann2005} when analyzing the \xmm\ data. The X-ray continuum at energies above 3 keV was shown to be too hard in comparison with the thermal bremsstrahlung of the jet matter, while the emission lines of hydrogen- and helium-like iron pointed to a plasma temperature $\sim10$--15  keV. To explain the nature of the additional component, \citep{Medvedev2010} proposed a scenario in which the hard X-ray emission results from the reflection of the emission of a hypothetical central
source from opaque walls of a supercritical disk funnel. Apart from the excess of hard X-ray emission, this model also explains the observed fluorescent Fe\,I  line at 6.4 keV. In that case, the additional component and the observed fluorescence are not associated directly with the jet emission, while the required luminosity of the hidden source is   $\sim 10^{40}$ erg\,s$^{-1}$, which, however, may well be the case if SS~433 is an ultraluminous X-ray source
viewed edge-on  \citep[see also ][]{Middleton2018}.  In our recent paper  \cite{Medvedev2018}  we proposed an alternative scenario, in which, on the contrary, the hard component is explained by the emission from the hot extension of the visible part of the jets that is partially absorbed as a result of its passage through the dense wind of a supercritical disk.

The unknown formation mechanism of the hard X-ray emission makes it difficult to analyze the X-ray emission from the jets, because the unknown
spectral shape of the additional component does not allow the jet parameters to be reliably determined by modeling the broadband spectrum. At the same time, for a better understanding of the nature of the hard component, it seems necessary to more accurately extract its emission from the source's total observed spectrum by determining the contribution of the jet emission from the soft spectral region. In this paper we perform a systematic analysis of the high resolution X-ray spectra obtained by the \chandra\ observatory to determine the parameters of the jets within the multitemperature model of their emission that self-consistently predicts the line and continuum spectrum \citep{KMS2016}. Using the soft 1--3 keV energy band, where the relative intensity of the hard component is low, while the energy resolution and sensitivity of the HETGS reach the highest values, we extract the contribution of the additional component from the total broadband spectrum and reach conclusions regarding the validity of the proposed emission model for the hot extension of the jets \citep{Medvedev2018}.

The paper has the following structure. The set of investigated observational data from high-resolution X-ray spectroscopy is described in Section~2. 
Section~3 gives a brief description of the \texttt{bjet} X-ray jet emission model described in detail in \cite{KMS2016}. In Section~4 we analyze simplified characteristics of the spectra and compare them with the predictions of the jet emission model and show that the spectral shape of the observed emission in the entire accessible range cannot be described exclusively by the jet emission model. In Section~5 we perform a detailed analysis of the spectrum below 3 keV, where the general emission characteristics are consistent with the predictions of the jet emission model and for which the best quality of spectroscopic data is available. In Section~6 the results obtained from the soft spectral band are compared with the broadband spectrum. In Section~7 we propose a physical model of the observed emission excess and fit the broadband spectrum by a combined model. In Section~8 we formulate our main conclusions.

\section{Data}
\label{s:data}
\begin{table*}
\begin{center}
\begin{tabular}{llcccccr}
\hline
\obs & Date &  JD 2'450'000$+$  &  Exposure,, &  Count rate, & $\phi$ &   $\psi$ & Bibliography  \\
 & & & ksec  & cts/s (MEG)& & & \\
\hline
106   &  1999-09-23 &  1445.01 &   28.7 & 1.59 & 0.64 &  0.92 & \cite{Marshall2002} \\
1020  &  2000-11-28 &  1877.06 &   22.7 & 0.58 & 0.67 &  0.58 & \cite{Lopez2006} \\
1019  &  2001-03-16 &  1985.43 &   23.4 & 1.58 & 0.95 &  0.25 & \cite{Lopez2006}, \\
&&&&&&&\cite{KMS2016}\\
1940  &  2001-05-08 &  2038.07 &   19.6 & 0.35 & 0.97 &  0.57 & \cite{Marshall2013a}\\
1941  &  2001-05-10 &  2039.93 &   18.5 & 0.22 & 0.12 &  0.58 & \cite{Marshall2013a}\\
1942  &  2001-05-12 &  2041.94 &   19.7 & 0.88 & 0.27 &  0.60 & \cite{Namiki2003}, \\
&&&&&&&\cite{Marshall2013a}\\
5512  &  2005-08-06 &  3588.97 &   19.7 & 2.31 & 0.52 &  0.13 & \cite{Marshall2013a}\\
5513  &  2005-08-12 &  3594.83 &   48.1 & 1.75 & 0.97 &  0.16 & \cite{Marshall2013a}\\
5514  &  2005-08-15 &  3597.93 &   73.1 & 2.08 & 0.21 &  0.18 & \cite{Marshall2013a}\\
6360  &  2005-08-17 &  3600.45 &   57.3 & 2.07 & 0.40 &  0.20 & \cite{Marshall2013a}\\
15781 &  2014-08-09 &  6878.98 &  138.2 & 0.33 & 0.01 &  0.40 & \\
\hline
\end{tabular}
\caption{\small
Log of \chandra/HETGS spectroscopic observations of SS~433. The columns present the observation identifiers
and dates, exposure times, count rates, orbital and precession phases, and bibliographic references to the previous papers
where the corresponding observations were used.} 
\label{tab:obs}
\end{center}
\end{table*}
 
 Table~\ref{tab:obs} provides general information including the observation identifiers, dates, exposure times, count rates, orbital and precession phases for the \chandra observations of SS~433 analyzed in this paper. The last column lists the references to the previous papers where the corresponding observations were analyzed. All observations were obtained using the ACIS (Advanced CCD Imaging Spectrometer) in combination with the sensitive HETGS (High-Energy Transmission Grating Spectrometer,  \citealt{Weisskopf2002,Canizares2005}). For each observation we combined the spectra taken in the $\pm 1$st diffraction orders, but the data from the high- and medium energy gratings (HEG and MEG) were analyzed separately due to a significant difference in their response functions.

\begin{figure}
\begin{center}
\hspace*{-0.5cm}\includegraphics[width=0.92\columnwidth]{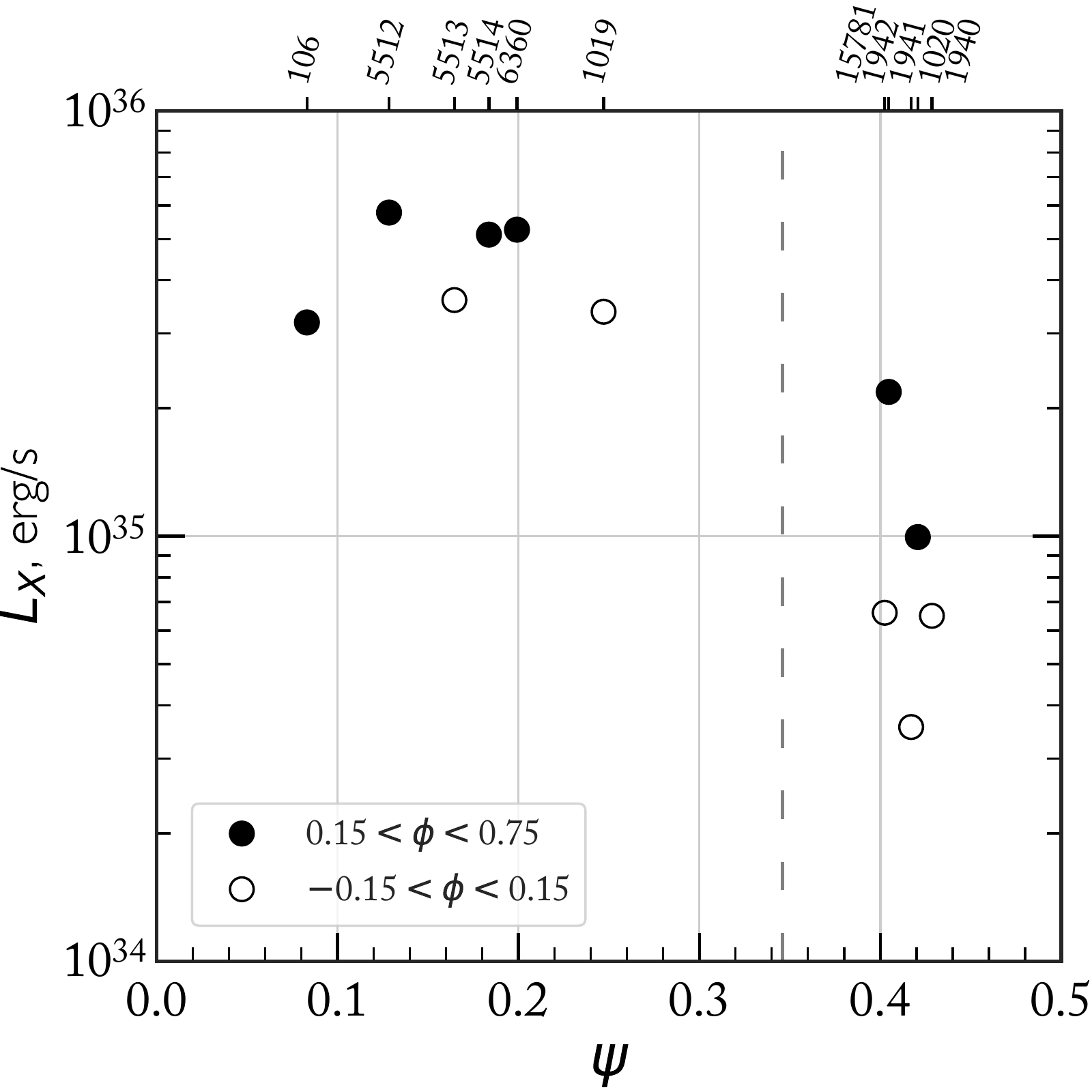}
\caption{\small
X-ray luminosity of SS~433 recorded by the Chandra MEG (0.4--5.0 keV) during the observations presented in Table~\ref{tab:obs} versus its precession phase $\psi$. The precession phase was calculated from the ephemeris of \protect\cite{Goranskij2011}; for the observations with $\psi>0.5$  the precession phase was mirrored according to the formula $\psi = 1 - \psi$. The X-ray luminosity was calculated for a distance to SS~433 taken to be 5 kpc.   The numbers at the top indicate  \obs\  for each observation. The measurement errors are comparable with the symbol size. The out-of-eclipse observations (orbital phases  $0.15<\phi<0.85$) are shown by the filled symbols.  The vertical dashed line marks the times at which the jet axis crosses the observer's plane of the sky (crossover times).} 
\label{fig:fluxes}
\end{center}
\end{figure}

The data were downloaded, prepared, and processed by means of the standard \texttt{TGCat} \citep{Huenemoerder2011}  and \texttt{CIAO 4.9} packages. 
Figure~\ref{fig:fluxes} shows the X-ray luminosity of SS 433 converted from the fluxes presented in the \texttt{TGCat}  flux properties table
 \citep[see.][]{Huenemoerder2011}  for the MEG energy range and a distance to the source taken to be 5 kpc.
Since the \texttt{TGCat} standard flux determination procedure does not involve modeling the source's spectrum, the quantities presented in Fig.~\ref{fig:fluxes} are designed only for a qualitative description of the SS~433 states.

The spectra were fitted by means of standard tools from the  {\sc XSPEC}  software package  \citep[version 12.10.1,][]{Arnaud1996} and the Python interface PyXSPEC (version 2.0.2) for the organization of a data analysis pipeline. Some fraction of the spectral channels have a low signal-to-noise ratio; in this case, applying the standard  $ \chi^2$-statistic can lead to biased estimates of the best-fit parameters \citep{Humphrey2009}.
Given that we are interested in analyzing the spectral lines of the jets, we prefer to avoid additional data binning due to the loss of information
inherent in such a procedure. Taking into account the low HETGS noise background level, instead we will use the \cite{Cash1979} statistic (C-statistic),
which is applied for data with Poissonian statistics and, at the same time, asymptotically tends to $ \chi^2 $ statistic in the limits of a large number of counts.
The spectra with a small number of counts can also be analyzed using statistical weighting \cite{Churazov1996}, as was demonstrated, for example, in
\cite{KMS2016}. The significance of the result obtained and the degree of degeneracy of individual model parameters were determined by the
Monte Carlo method based on a scheme of Markov chains. The Metropolis-Hastings algorithm \citep{Hastings1970}  was chosen as the scheme of a Markov chain. The errors for the best-fit parameters given in this paper correspond to the intervals between the 5 and 95\% quantiles (90\% significance).

\section{Multitemperature thermal jet emission model}
\label{s:bjet}
The multitemperature jet emission model with adiabatic expansion and gas energy losses by radiation was calculated numerically by \cite{Brinkmann1988}
 and analytically by \cite{Koval1989} (without allowance for the line emission when calculating the spectrum and the gas cooling function). In subsequent papers  \citep{Medvedev2010, KS2012, KMS2016} 
 the calculations have already been based on modern atomic databases, which allow the emission in lines and continuum to be taken into account. The model from  \cite{KMS2016}
looks particularly convenient for observational data analysis owing to the tabular presentation of the computed models on the physically motivated domain of the parameter space (below in the text referred to as
the \texttt{bjet} model). In this paper the \texttt{bjet} model will be used as the main tool for analyzing the observational data; the basic principles of the model are described below (for a detailed description, see \citealt{KMS2016}).

The relativistic jet is treated as an axisymmetric ballistic flow of baryonic matter (gas) moving with a constant, mildly relativistic speed $ \beta = v/c $  perpendicularly to the accretion disk plane. The degree of gas collimation is specified by the flow cone half-opening angle  $\Theta \sim 0.01$  rad. For such a formulation of the problem the physical conditions along the entire jet are functions of only one coordinate along the axis of symmetry. 
The basic equation defining the gas temperature profile along the jet is the thermal balance equation:
\begin{equation}
\label{eq:cooling}
   \frac{dT}{dr}=-2\left(\gamma-1 \right)  \frac{T}{r}-\frac{2n_{e} n_{i}}{3\left(n_{e}+n_{i}\right)} \frac{\Lambda_{Z} \left( T \right)}{\beta c},
\end{equation}
where $r$ is the distance measured from the cone vertex along the jet axis, $n_e(r)$ and $ n_i(r)$ are the electron and ion densities, T(r) is the gas temperature, and
$\gamma=5/3$ is the adiabatic index. The first and second terms on the right-hand side of the equation correspond
to cooling due to adiabatic expansion and gas energy losses by radiation, respectively. The integral gas emissivity $\Lambda_{Z} \left(T\right)=\int\epsilon_Z(E,T)dE$  is calculated
in the regime of a hot optically thin plasma in collisional ionization equilibrium (CIE) based on  \texttt{AtomDB/APEC}\footnote{\url{http://www.atomdb.org}} \citep[version 3.0.9, ][]{Foster2012}. Thus, the gas energy losses by radiation are calculated self-consistently, which allows the emission measure distribution to be reproduced with a high accuracy in the range of temperatures corresponding to both continuum and line emissions \citep{KMS2016}. The total model spectrum is determined by adding the contributions of thin single-temperature transverse layers along the jet. The input parameters of the  \texttt{bjet}\footnote{We use its version adapted for analyzing the spectra of SS 433; for more details, see \cite{KMS2016}.} are the kinetic luminosity $L_k$, the gas temperature at the jet base $T_0$ (the jet region closest to the compact object directly visible to an observer is called the base), the transverse optical depth for electron scattering at the jet base $\tau_{e0}$, and the heavy-element abundance $Z_i$ relative
to the solar chemical composition \citep{Anders1989}  (in this paper we will decouple the abundances of various elements when fitting the data; therefore, we prepared a similar model with free abundance parameters\footnote{The model is publicly accessible and can be downloaded from \url{ftp://hea.iki.rssi.ru/medvedev/bjet}}). 
The shape of the model spectrum is determined by the differential emission measure distribution along the jet and depends mainly on the parameter $\alpha$ (but also on $T_0$ and $Z_i$), which is the ratio of the energy losses by radiation to the adiabatic cooling at the jet base:
\begin{equation}
\alpha = \frac{2}{3}\frac{\tau_{e0}}{\Theta\beta}\frac{\Lambda_Z(T_0)}{\sigma_e c T_0} \frac{X}{1+X},
\end{equation}
where $\Lambda_Z(T_0)$ is the plasma emissivity, $X = n_i/n_e \approx 0.91$ is the ratio of the ion and electron densities, and $\sigma_e=6.65 \times 10^{-25}$ cm$^2$ is the Thomson cross section.
If the energy losses by radiation are assumed to be due to only the bremsstrahlung of hydrogen and helium, then we can obtain a simple estimate, $\alpha \approx  4.42~\tau_{e0} \times~\left(\frac{10~{keV}}{T_{0}}\right)^{1/2}$. The gas cooling is determined by the  adiabatic jet expansion for  $ \alpha \ll 1$ and by the gas energy losses by radiation for $ \alpha \gg 1$ (for more details, see \citealt{KMS2016}).

\section{Basic characteristics of broadband spectra} 
\label{sec:hard_component} 

\begin{figure}
\centering
\hspace*{-0.5cm}\includegraphics[width=1\columnwidth]{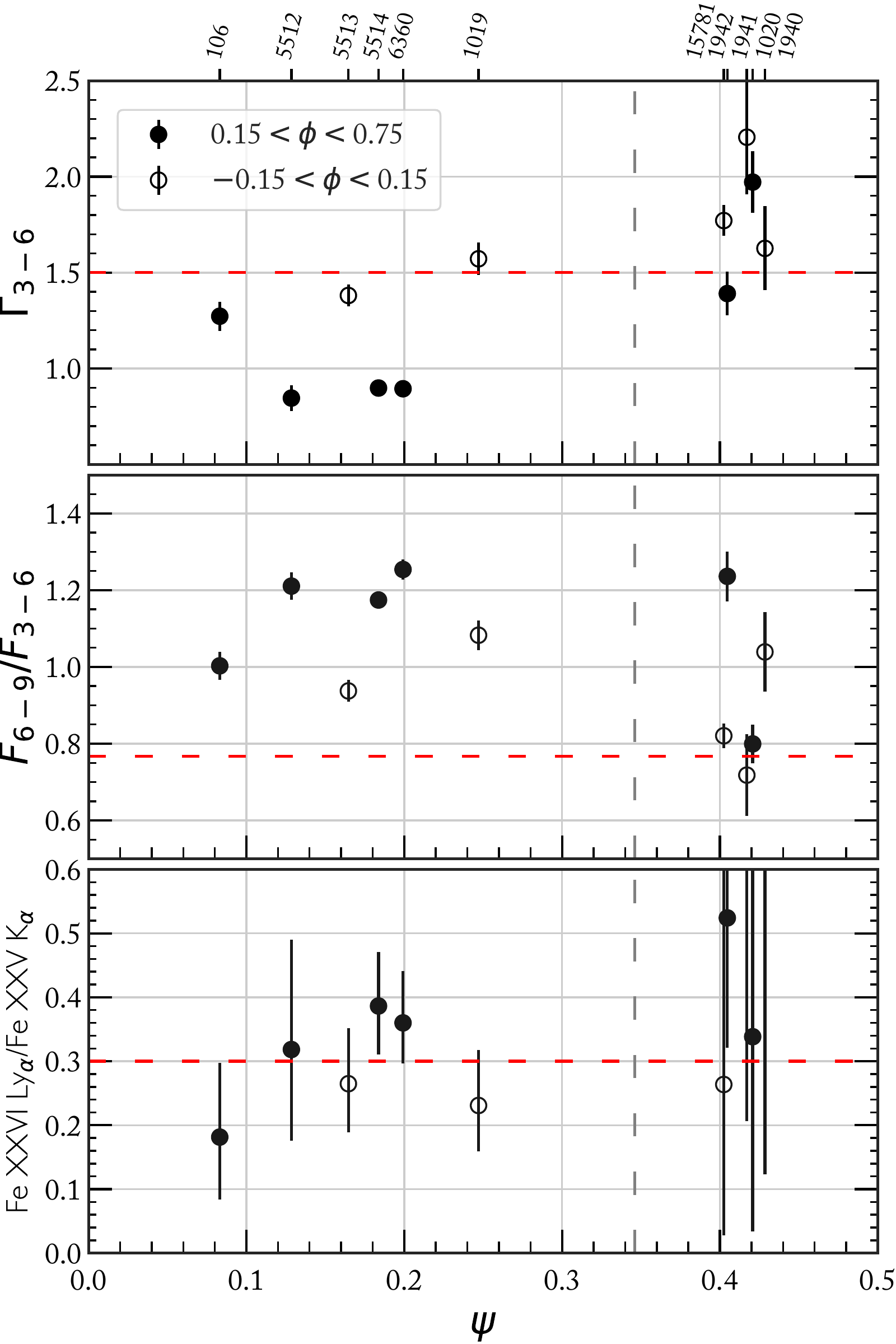}
\caption{\small 
Basic characteristics of the SS~433 spectra for the \chandra/HETGS observations versus precession phase $\psi$, for the observations with phases $\psi>0.5$  the precession phase is mirrored. The upper panel shows the
spectral power-law index in the 3--6 keV energy band. The horizontal dashed line indicates  $\Gamma=1.5$. 
The middle panel shows the ratio of the fluxes in the 6--9 and 3--6 keV energy bands found by fitting the spectra with a power law and a sum of six Gaussians in the 3--9 keV band. The horizontal dashed line indicates a flux ratio of $\approx 0.75$ for the power-law spectrum with $\Gamma=1.5$. The lower panel shows the ratio of the fluxes in the \feLy\ and \feK. The horizontal dashed line
indicates 0.3 corresponding to a jet base temperature of $\approx 12$  keV (see Fig.~7 in \citealt{KMS2016}). All quantities were
corrected for absorption for an equivalent hydrogen column density of  $1.2 \times 10^{22}$  atoms cm$^{-2}$. The vertical line marks the times at which the jet axis crosses the observer's plane of the sky (crossover).} 
\label{fig:observables}
\end{figure}

The multitemperature jet model predicts a powerlaw spectrum in the 3--6 keV energy band, which is a sum of plasma bremsstrahlung with different
temperatures. The spectral slope turns out to be sensitive to the temperature of the visible jet base in view of the decrease in differential emission measure
with decreasing gas temperature along the jet and depends weakly on other model parameters in view of the absence of intense emission lines in this energy band. In contrast, the 6--9 keV X-ray flux depends largely on the flux in iron and nickel lines, which
makes the 6--9 to 3--6 keV flux ratio sensitive to the gas metallicity in the jets, because the continuum in this energy band predicted in the model is formed
mainly through the bremsstrahlung of hydrogen and helium. Measurements of the fluxes in hydrogen- and helium-like iron lines, the most intense lines of the jets in the X-ray wavelength range, serve as another tool for diagnosing the jet base temperature (for a description of the jet parameter estimation method shown for \obs\ 1019 as an example, see  \citealt{KMS2016}).

As a preliminary estimate of the jet parameters we obtain the described spectral characteristics for the data sample listed in Table~\ref{tab:obs}. For this purpose, we describe the observed spectra in the 3--6 keV energy band by a power law with absorption. The fitting results are shown on the upper panel in Fig.~2. Next, by extending the energy band to 9 keV, we model the main observed lines by a sum of Gaussians with linked Doppler shifts corresponding to the line-of-sight velocities of the two jets. In this energy band the most intense lines of the jets are the \feK, \feLy\ and \niK\ lines. The centroid of the unshifted \niK\ line ($E_0\sim7.8$ keV, the weighted mean centroid of the triplet) lies at the sensitivity edge of the first HEG diffraction order and, therefore, the line is detected with confidence only for the jet with a positive line-of-sight velocity. Apart from the relativistic lines, the fluorescent \feI\ line at 6.4 keV with an equivalent width of $\approx 50$--60 eV is detected reliably; we will describe this line by a Gaussian with a fixed centroid. The ratios of the fluxes in the 6--9 and 3--6 keV bands and the ratios of the fluxes in the  \feLy\ and \feK lines found from our fitting are shown on the middle and lower panels of Fig.~2, respectively.

The observations under consideration can be divided by their spectral characteristics into three groups. The first group of observations (\obs\  5512, 5514, and 6360) is characterized by a hard continuum with a photon index $\Gamma\sim1$ and the greatest relative
contribution of the emission lines. In contrast, for the second group of observations (\obs\   1940, 1941, 1942, 1020, and 15781) the continuum is described by a power-law slope $\Gamma\sim2$, while the jet lines in the 6--9 keV band are poorly detected. The observations from the second group correspond to the precession phases when the accretion disk of SS~433 is viewed nearly edge-on, which may point to a vast structure of the accretion disk wind eclipsing the hot jet base. The third, ``intermediate'', group of observations
(\obs\ 106, 5513, and 1019) corresponds mainly to orbital phases  $\phi<0.15$ and $\phi>0.85$, during which
the emission from the hot accretion disk regions and the jets near the compact object could be partially hidden from the observer by the companion star. At the same time, the spectral characteristics of \obs\ 106 and 1020 clearly point to
an intrinsic variability of the central X-ray source in SS~433: for the first observation the continuum in
the 3--6 keV band is appreciably ``softer'' than that in other out-of-eclipse observations in this range of
precession phases, while during the second observation the jet emission lines were almost completely
absent in both the soft part of the spectrum 1--3 keV \citep{Lopez2006} and the 6--9 keV energy band (see
the middle panel in Fig.~\ref{fig:observables}). While comparing Figs.~\ref{fig:fluxes} and~\ref{fig:observables}, we can note an inverse correlation between the X-ray flux recorded by the MEG and the power-law slope in the 3--6 keV energy band.

It can be seen from Fig.~2 that the power-law slope
of the spectra for the first group of observations is
considerably smaller than $\Gamma=1.5$, the limiting value
in the jet emission model for a base temperature of  $\approx 40$ keV. At the same time, the ratios of the fluxes
in the \feLy\ and  \feK\ lines point to a moderate gas temperature near the jet base, $\approx 10$ keV (cf. with Fig. 7 from  \citealt{KMS2016}).
Such a contradiction was also pointed out based on \xmm/EPIC-PN observations \citep{Brinkmann2005}. For the observations from the second
group the \feLy\ and \feK\ lines are barely detected and, therefore, the flux ratios found for them turn out to be poorly constrained.

\section{Diagnostics of the jets from the soft part of X-ray spectra}
\label{s:soft}
The unknown spectral shape of the additional component introduces great uncertainties in describing the Chandra broadband spectra of SS~433 by the jet emission model. Therefore, it seems important to attempt to separate the X-ray emission produced in the jets from the emission of the additional component, which must allow both the jet parameters and the formation mechanisms of the observed hard continuum to be determined more accurately. For this purpose, the 1--3 keV energy band, where the bulk of the X-ray lines are emitted and the contribution of the additional component is expected to be comparatively
small, looks most suitable. The high efficiency of the HETGS gratings in this energy band (the best energy resolution and the highest sensitivity) allows the intensities and positions of the lines of the approaching and receding jets to be measured separately, which makes it possible to study their motion characteristics, geometry, and the emission measure distribution in the gas along the jets.

The precession phases near  $\psi=0$, when the lines of the two jets are separated better and the inclination of the jets to the line of sight changes slowly, appear to be most optimal for determining the jet parameters from the soft part of the X-ray spectra. The
\obs\  5512 and 6360 observations, during which no abrupt changes in the X-ray flux and line positions
were observed, look most ``stable''. During the \textit{VLBA} and \chandra\ \citep{Marshall2013a} observations corresponding
to \obs\ 5513 and 5514, an abrupt shift in the positions of the jet emission lines was detected over the exposure time, which was accompanied by the ejection of a portion of the gas in the radio jets. Such ``jitter'' of the X-ray jets was also detected during other observations: during the joint observational
 campaign of 2006 in the X-ray (\textit{Suzaku} observatory) and optical (6-m BTA telescope) wavelength ranges \citep{Kubota2010} and during the recent \xmm\ observations with a total exposure time of  $\approx 120$  ks \citep{Medvedev2018}. In this paper we investigate the influence of such a spectral variability of the X-ray source in SS~433 on the results of our data analysis by dividing each observation into two equal parts, after which each part is analyzed by the same method as are the total-exposure data.

The data were analyzed by means of standard tools from the {\sc XSPEC} software package \citep[version 12.10.1,][]{Arnaud1996}. Following  \cite{KMS2016}, the contribution of the two oppositely directed jets was calculated using a superposition of two \texttt{bjet}  models (the \texttt{bjet}  model is described in Section~\ref{s:bjet}). The Doppler shift of the jet spectral lines was specified by the \texttt{zashift}  convolution model (the relativistic boosting effect is also taken into account in the model). The line profile and width were specified by smoothing the model spectrum with a Gaussian using the \texttt{gsmooth} tool with a power-law index of 1, in accordance with the model of line broadening as a result of ballistic jet gas expansion. The interstellar and internal photoabsorption of the X-ray emission from SS~433 was taken into account using the multiplicative  \texttt{phabs} model with an abundance of the absorbing material specified in accordance with   \cite{Asplund2009}\footnote{We assume that the absorption of the X-ray emission from SS~433 can be partly attributable to photoabsorption in the gas inside the system and, therefore, the set of abundances for interstellar matter from \cite{Wilms2000} was not used.}  We performed tests using other absorption models, namely \texttt{phabs},  \texttt{TBabs} and \texttt{TBgas},
and tests of these models in combination with the abundance from  \cite{Wilms2000}. The equivalent hydrogen column densities  ($N_H$) found by analyzing the data with the  \texttt{phabs},  \texttt{TBabs} and \texttt{TBgas} absorption models for two sets of abundances of the absorbing
material agree within 5\%, while $N_H$  for the \texttt{wabs} model is considerably lower, on average, by 25\%. The combined model used to describe the observed spectra at energies up to 3~keV can be arbitrarily written as follows:

\begin{equation}
\begin{split}
\textsf{Model} = phabs* (gsmooth * (zashift_{b}*jet_{b}+ \\ constant*zashift_{r}*jet_r )),
\label{eq:model}
\end{split}
\end{equation}
where the factor $constant$ allows the additional attenuation of the red-jet emission by a constant factor in the entire energy range to be taken into account. We will assume both jets to be identical in terms of both geometrical characteristics and physical properties of the gas in them. Therefore, the number of free model parameters is reduced by coupling the kinematic luminosity, the base temperature, the elemental abundances, and the line widths for the two jets.

\subsection{Fitting results}
\label{s:fitting}
The initial Doppler-shift parameters and the red jet suppression factor  $constant$  were specified in accordance with the quantities found by analyzing
the spectral characteristics in the 6--9 keV band (see Section~\ref{sec:hard_component}). As an initial guess for the line width (Gaussian root-mean-square (rms) width) we used $\Sigma(E) \approx 25~eV  \times (E/6~keV)$ corresponding to the opening angle and velocity of the jets specified in the \texttt{bjet} model. The initial chemical composition of the jets was specified to be the solar one (\citealt{Anders1989};  for a discussion, see below), except
for the abundance of nickel whose excess has been repeatedly confirmed both from the soft part of the spectrum and in the 6--9 keV energy band; we set the initial nickel abundance equal to $Z_{Ni} = 8$ \citep{Brinkmann2005, Medvedev2010, KMS2016, Medvedev2018}.
In the first fitting step, we searched for the jet line positions in the spectrum. For this purpose, all parameters, except for
the Doppler line shifts, the spectrum normalization, and the temperature, are fixed. The spectrum is fitted
in the 0.8--3 and 1--3 keV energy bands for the MEG and HEG, respectively. In the second step, the line positions are fixed and the absorption is determined from the data. Next, the best-fit model is determined for the Ne, Na, Al, Mg, S, Fe, and Ni abundances, the jet line widths, and the factor $constant$ at a fixed solar Si abundance.

Figures~\ref{fig:fit6360} and~\ref{fig:fit5513}  show the best-fit models found from the soft part of the X-ray spectra for  \obs\  6360, 5512, 5513, and 1019.  \obs\  6360 and 5512 refer to the ``hard'' out-of-eclipse group of spectra (see Section~\ref{sec:hard_component}). 
 \obs\  5513 and \obs\  1019 correspond to the system's orbital phases near the maximum eclipse depth (see Table~\ref{tab:obs}). It is clearly seen from a comparison of \obs\ 6360, 5512, and 5513 that the soft part of the spectra is hardly eclipsed (see also \citealt{Marshall2013a}) at the times of minimum inclination of the disk axis and the jets to the line of sight (precession phases $\psi < 0.2$). The \obs\  1019 observation shows that the spectra are difficult to fit during precession phases close to the crossover, when the lines of the two jets are closely
spaced relative to one another. The numerous Ne, Na, Fe, and Ni lines at energies $<1.5$ keV are seen to blend together, what creates difficulties in determining the abundances of these elements, the line widths and centroid positions. Moreover, for other
observations near the crossover (the ``soft'' group of spectra), apart from the difficulty of separating the jet lines, there are problems related to the poor data quality (low signal-to-noise ratio), which, in turn, is a consequence of the low count rate during the corresponding
observations (see Table~\ref{tab:obs}). For the ``hard'' group of spectra the model specified by Eq.~\ref{eq:model} gives a good description of the continuum shape and normalization up to energy of $\approx 2.5$ keV; at higher energies an excess of the flux relative to that predicted in the model is observed probably due to an increase in the relative contribution of the additional component. The presented best-fit models have the following values of the C-statistic divided by the number of degrees of freedom: $12083/11147=1.08, 11063.48/11147=0.99, 11890/11147=1.07$ and $11390.85/11147=1.02$.

A great number of lines originating in the jets, including the neon, sodium, magnesium, aluminum, silicon, sulfur, iron, and nickel lines, are clearly detected in the
part of the spectrum under consideration. The high resolution of the spectrum allows the lines of different ionization states of the elements to be separated. All the lines detected with a good significance belong to the relativistic jets. Only the weak line with the centroid at 2 keV ($E_0\approx1.92$  keV for $z= -0.044$) between the \siK\ and \siLy\ lines, which is clearly seen during the \obs\ 6360 observation, constitutes an exception. No such line was detected for other observations, which may be the result of blending with the more intense jet lines falling at 2 keV. The largest deviation from the model predictions is observed for the hydrogen-like \siLy\ doublet lines from the \obs\  6360 data, while other silicon lines are described satisfactorily. Our data analysis revealed no such problems for the analogous sulfur and magnesium lines (see Figs.~\ref{fig:fit6360} and~\ref{fig:fit5513}).

\subsection{Temperature and composition of the X-ray jets}
\label{s:composition}
The emission lines observed in the soft part of the standard X-ray band originate predominantly in jet regions with temperatures $<3$ keV, at which the maximum plasma emissivity in a specified line is reached. For example, for the S\,XVI\,Ly$\alpha$ ($E_0 = 2.62$ keV) line the emissivity peak is reached at a plasma temperature of  $\approx 2.2$ keV. At the same time, the observed ratio of the fluxes in the \feLy\ and \feK\  lines (Fig.~\ref{fig:observables}) clearly points to a jet base temperature $T_0\gtrsim 10$  keV. In that case, the hot gas near the jet base radiates its energy predominantly through the bremsstrahlung of hydrogen and helium and, thus, makes a significant contribution to the continuum flux in the soft part of the spectrum, while the line emission originates in considerably colder parts of the jets. Therefore, the observed equivalent width of a specified line for an arbitrary element can be obtained for various combinations of its abundance and the jet base temperature by varying which we can regulate the contribution of the hot-base emission in the energy band of the line under consideration. The base temperature $T_0$, along with other model parameters, specifies the plasma differential emission measure distribution along the jet, thereby changing the relative contribution of the lines of various elements. However, in the case of quasi-adiabatic gas cooling near the base ($\alpha\ll 1$), which is assumed based on the sizes of the X-ray jets in SS~433 and their observed X-ray luminosity (see \citealt{KMS2016}), the differential emission measure distribution takes an invariant form relative to a variation in model parameters:
\begin{equation}
DEM(T/T_0) = \frac{dEM}{d\ln (T/T_0)}=\frac{n_e n_i~dV}{d\ln T/T_0} \propto \left(\frac{T}{T_0}\right)^{3/4}. 
\end{equation}
In that case, each transverse single-temperature gas layer of the jet makes a constant relative contribution to the total emergent model spectrum. Therefore, within the formalism under consideration the base temperature turns out to be degenerate relative to the absolute normalization of the elemental abundances when fitting the data in the 1--3 keV energy band. Apart from the continuum level, the base temperature, obviously, affects the slope of the multitemperature bremsstrahlung spectrum for the jets, which, however, exerts a weak effect on the spectrum in the narrow 1--3 keV energy band. In practice this effect is difficult to feel, because near 3 keV the situation is complicated by a significant contribution of the hard component, which, in addition, changes considerably from observation to observation.

Nevertheless, the jet base temperature can be determined by assuming the chemical composition of the gas in the jets to be close to the solar one, which allows the model flux in emission lines to be calculated. We use this assumption by fixing the silicon abundance in the last data fitting step in accordance with the solar value from \cite{Anders1989}. At the same time, the abundances of the remaining elements remain free model parameters, which allows the ``relative'' elemental abundances and the jet base temperature corresponding to the observed line equivalent widths to be calculated, provided that the silicon abundance is solar. As a normalization we chose the silicon abundance in view of the largest contribution from the lines of this element to the 1--3 keV flux ($\approx 5$--7\%). Besides, the flux in silicon lines is clearly detected owing to the intense Si\,XIV\,Ly$_{\alpha}$ and Si\,XIII\,K$_{\alpha}$ lines, which rarely overlap with the lines of the opposite jet.

\begin{figure}
\centering
\includegraphics[width=1\columnwidth]{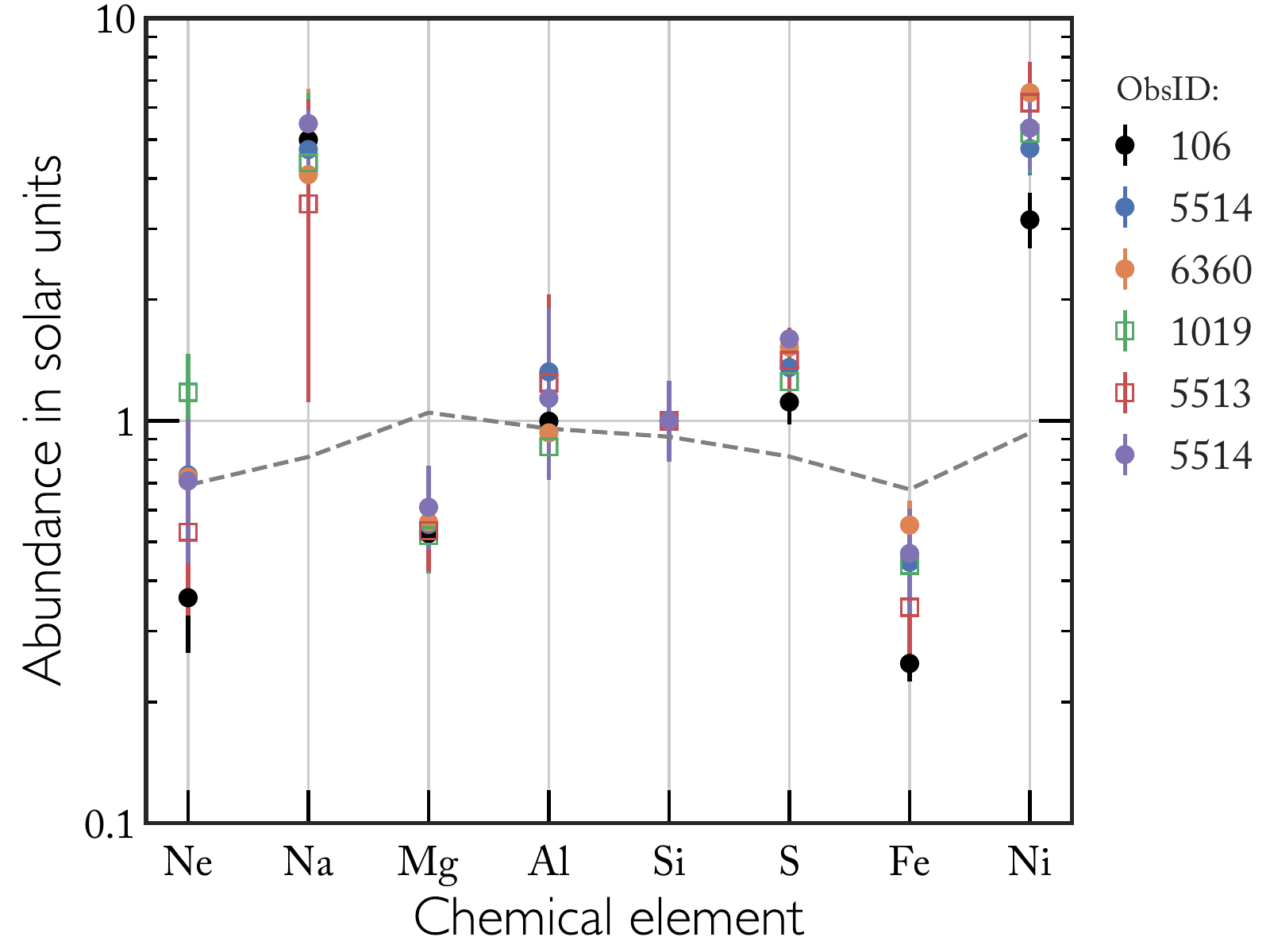}
\caption{\small  
The gas composition in the jets of SS~433 found from the soft part (1--3 keV) of the  \chandra/HETGS X-ray spectra. The silicon abundance is taken to be the solar one \citep{Anders1989}.  The derived neon (Ne), sodium (Na), magnesium (Mg), silicon (Si), sulfur (S), iron (Fe), and nickel (Ni) abundances are shown for the sample of observations from Table~\ref{tab:obs}. The dashed line indicates the set of elemental abundances corresponding to the solar chemical composition from \protect\cite{Asplund2009}.} 
\label{fig:composition}
\end{figure}

Figure~\ref{fig:composition} shows the elemental abundances in the jet gas in units of the solar abundance \citep{Anders1989} found from the 1--3 keV emission lines. The dashed line indicates the set of solar elemental abundances from \cite{Asplund2009}. For \obs\ 5513 and 5514 the measurements were made from the spectra obtained after the division of the total exposure into two parts; the values averaged over the two parts are given. We do not provide the measurements for  \obs\  1020, 1940, 1941, 1942, and 15781, due to the poor quality of the detection of jet emission lines. The  \obs\  106 observation, for which the spectrum at energies $<1.5$  keV is strongly absorbed and has poor statistics and, as a result, the abundances of many elements turn out to be shifted toward lower values, stands out among the gas composition measurements.

\begin{figure}
\centering
\includegraphics[width=0.9\columnwidth]{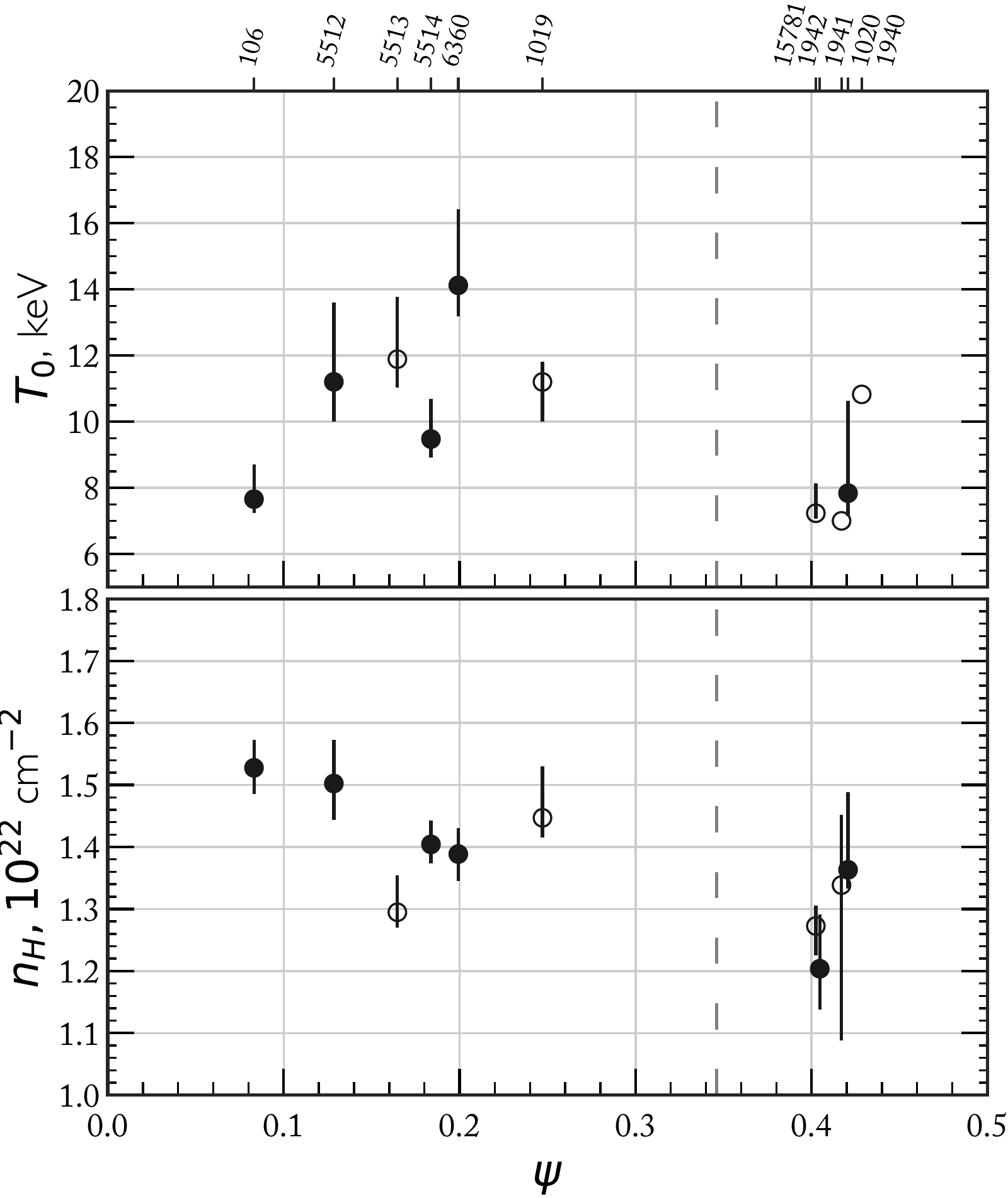}
\caption{\small  
Parameters of the best fit to the \chandra/HETGS data in the 1--3 keV energy band by the  \texttt{bjet} thermal baryonic jet
emission model at a fixed solar silicon abundance and free abundance parameters of other elements: Ne, Na, Mg, Al, S, Fe,
and Ni (see Fig.~\ref{fig:composition}). The system's precession phase  $\psi$ is along the horizontal axis; for the observations with phases $\psi>0.5$ the
precession phase is mirrored. The upper panel shows the jet base temperature for the observations from Table~\ref{tab:obs};
the lower panel shows the hydrogen column density for the \texttt{phabs} absorption model. The vertical dashed line indicates the
crossover position.
} 
\label{fig:T0}
\end{figure}

The sensitivity of the model to the iron abundance is determined mainly by the large number of lines at energies $<1.5$ keV corresponding to transitions
in Fe\,XVII--XXIV ions; the total contribution of the iron lines to the flux in the 1--3 keV energy band is comparable to the contribution of the silicon
lines. Note that the abundances of the elements with lines predominantly in the soft part of the spectrum ($<1.5$) depend most strongly on the adopted
hydrogen column density in the absorption model. As can be seen from Fig.~\ref{fig:composition}, the low relative iron
abundance in comparison with that from \cite{Anders1989} agrees much better with the more recent iron measurements in the solar photosphere from \cite{Asplund2009}. The neon abundance is reliably measured from Ne X lines; in particular, the most intense of them corresponding to the K$_{\alpha}$ transition and originating in the coldest regions of the X-ray jets (the temperature at the peak luminosity is $\approx 0.5$ keV) is reliably detected in the spectra under consideration. The value found also corresponds to the solar chemical composition. The sulfur abundance is determined slightly more poorly, because in the region of the most intense lines of this element the contribution of the additional component is significant, as a result of which the abundance being determined is shifted toward larger values. Taking this fact into account, we also conclude that the S abundance is nearly solar. The relative magnesium abundance is reliably determined from the intense Mg\,XII\,Ly$_{\alpha}$ line. The Mg\,XI\,K$_\alpha$ triplet is also detected in the best-quality spectra (\obs\ 6360 and 5512); the derived abundance turns out to be approximately half the solar value. The abundance of sodium, whose emission lines make the smallest contribution to the flux in the energy band under consideration ($<1\%$), is determined most poorly. The derived high sodium abundance results mainly from the residuals near the unshifted energy $E_0 \approx 1.275$ keV, where a large number of more intense Fe\,XXI, Ne\,X and Ni\,XIX-Ni\,XXV lines are emitted in addition to the Na\,X\,K$_\alpha$  line; therefore, the value obtained may be grossly overestimated. Remarkably, such a residual between the model and data is observed for the entire sample of observations under consideration, including the \obs\ 1019 observation near the crossover. The greatest deviation from the solar chemical composition was found for nickel. Intense nickel lines are reliably detected at energies below 1.5 keV. The derived nickel overabundance recalculated to $Z_{Ni}/Z_{Fe}\sim 10$, on the whole, is consistent with the flux measurements
in the  \niK\ and \niLy\ lines relative to the  \feK\ and  \feLy\  lines in the 6--9 keV energy band based on \xmm\ data \citep{Brinkmann2005, Medvedev2010, Medvedev2018}.

\begin{figure*}
\centering
\includegraphics[width=1\textwidth]{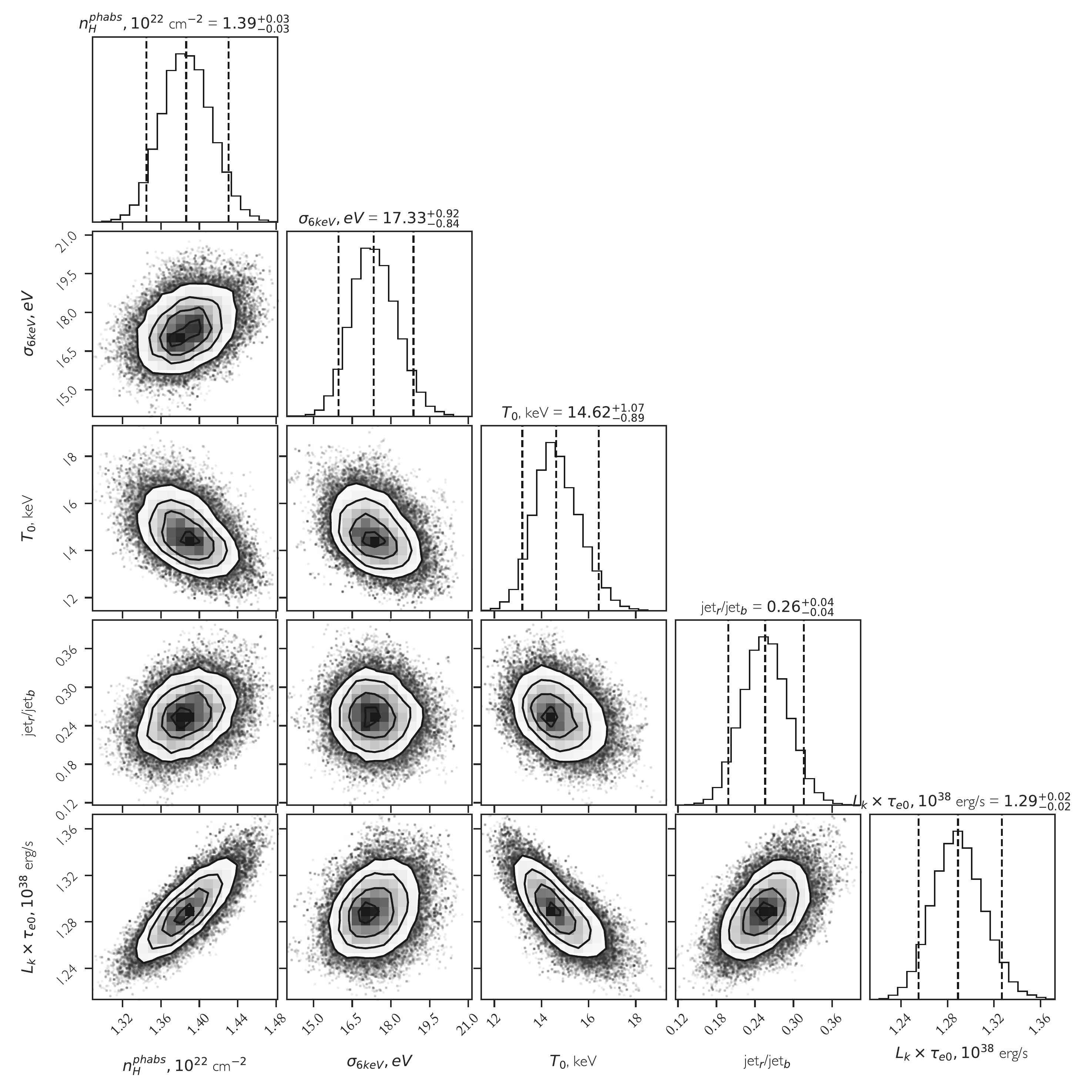}
\caption{\small 
One- and two-dimensional marginalized posterior distributions of parameters of the baryonic jet emission model (see
Eq.~\ref{eq:model}) for fitting the \obs\ 6360 data in the 1--3 keV energy band. The distributions are shown for the following parameters:
$n^{phabs}_H$ is the hydrogen column density in the \texttt{phabs}, photoabsorption model, $T_0$ is the jet base temperature (the temperatures
were fixed to be equal for the two jets), $\sigma_{6keV}$ is the rms width of the Gaussian jet line profile at 6 keV, and $L_k \times \tau_{e0}$ is the product of the kinetic luminosity by the optical depth for electron scattering at the jet base.$L_k \times \tau_{e0}$ was calculated by assuming the distance to SS~433 to be 5 kpc; $jet_r/jet_b$ corresponds to the red-jet suppression factor $constant$. The silicon abundance was fixed to be the solar one, while the abundances of the remaining elements emitting intense lines in the energy band under consideration were determined from our fitting. The 5\%, 50\%, and 95\% quantiles are indicated by the vertical dashed lines.
  } 
\label{fig:chains6360}
\end{figure*}

\begin{figure*}
\centering
\includegraphics[width=1\textwidth]{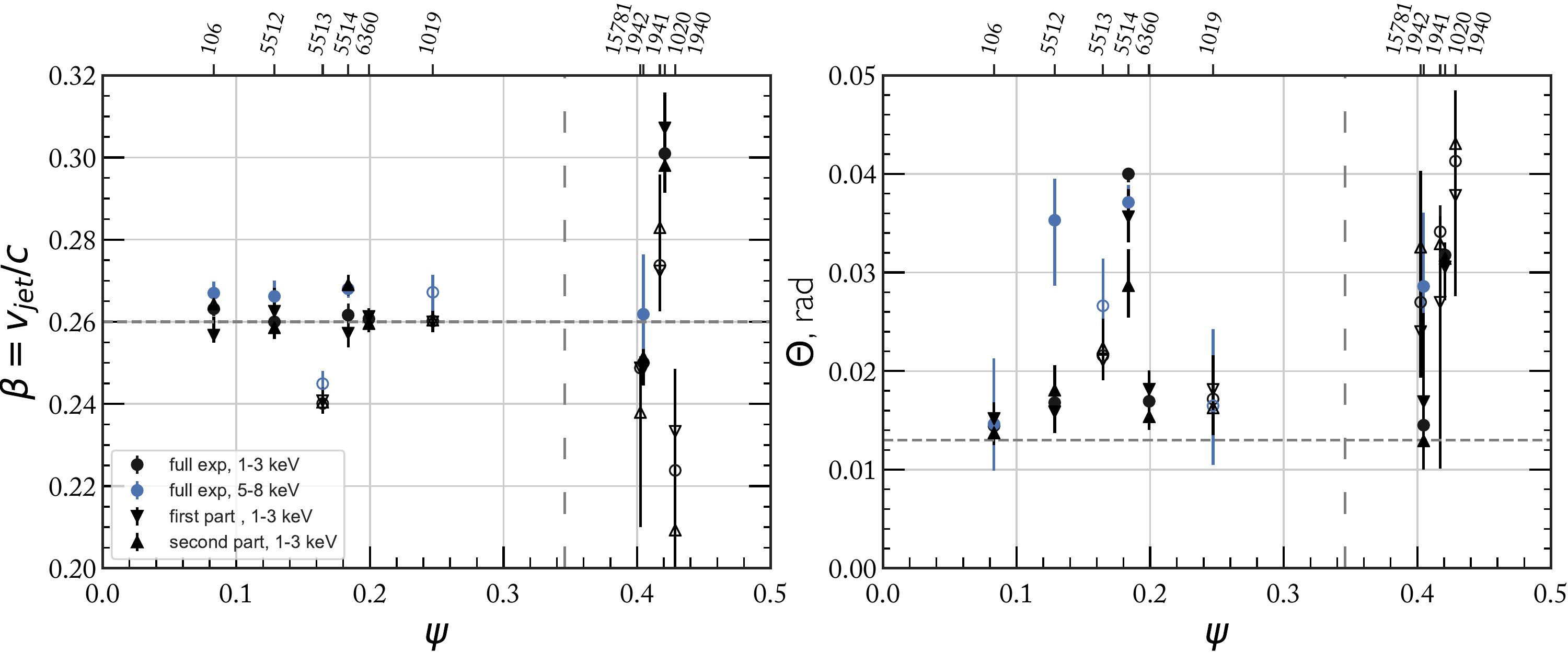}
\caption{\small 
Velocity, $\beta$ (a),and cone half-opening angle, $\Theta$ (b), of the SS~433 jets from the measurements of the
line positions and widths in the soft (1--3 keV, black symbols) and hard (6--9 keV, blue symbols) parts of the \chandra\ X-ray
spectra versus precession phase of the system $\psi$; for the observations with phases $\psi>0.5$ the precession phase is mirrored. The triangles indicate the measurements in the soft part of the spectrum when the exposure is divided into two parts
(the first part --- the downward-directed triangles). The horizontal dashed line on the left and right panels indicates $\beta=0.26$
and $\Theta=0.013$, respectively. The vertical dashed lines indicate the crossover positions.
} 
\label{fig:beta}
\end{figure*}

Figure~\ref{fig:T0} shows the derived jet base temperature that provides the observed line equivalent widths for the elemental abundances described above. The jet base temperature is seen to be virtually independent of the orbital eclipses, which is also clear from a comparison of the lines in the soft part of the spectrum in Figs. ~\ref{fig:fit6360} and \ref{fig:fit5513}. For most observations near the crossover the jet lines are poorly detected and, therefore, as in the case of elemental abundances, the base temperature is poorly constrained from the data. The model parameters are constrained most accurately from  \obs\  106, 5512, 5513, 5514, 6360, and 1019. Figure 5 presents the probability density distributions of the model parameters for  \obs\  6360. It can be seen from the two dimensional distributions that an increase in the base temperature corresponds to a decrease in the hydrogen column density. Nevertheless, provided that the chemical composition of the jets is nearly solar, all model parameters are satisfactorily constrained from the data.

\subsection{Jet opening angle and velocity}
The motion of jet lines in the spectrum of SS~433 results from a combination of the longitudinal and transverse Doppler effects. The transverse Doppler effect is clearly observed at the times at which the jets cross the plane of the sky (crossover times,  $\psi=0.35$ and $\psi=0.65$), when the Doppler shifts of the jet lines coincide and are equal to $v_r^{+,-}/c = \gamma -1 $, where  $\gamma=(1-\beta^2)^{-1/2}$ is the Lorentz factor and $\beta = v/c$ is the gas velocity in the jets in units of the speed of light. Measurements of the line shift allow the velocity of the jets and their geometry, the inclination to the line of sight, and the opening angle of the jets to be directly determined. Denote the Doppler shift of the jet that approaches the observer for most of the period by $z_b$, the Dopller shift of the opposite jet by $z_r$, and their mean by$z_0=(z_r+z_b)/2$. Then, assuming a perfect symmetry and the same jet velocity, we then obtain  \citep{KMS2016}:
\begin{equation}
\beta=\sqrt{1-\frac{1}{(1+z_0)^2}}=\sqrt{2z_0}\left(1-\frac{3}{4}z_0+\mathcal{O}(z_0^2)\right).
\label{eq:zbeta}
\end{equation}
Similarly, we can find the angle $\phi$ between the jet axis and the line of sight:
\begin{equation}
z_b=\gamma\left(1-\beta\cos\phi\right)-1,~~ z_r=\gamma\left(1+\beta\cos\phi\right)-1, 
\label{eq:zbzr}
\end{equation}
i.e.,
\begin{equation}
\cos\phi=\frac{1}{\beta}\left(1-\frac{1+z_b}{\gamma}\right)=\frac{1}{\beta}\left(\frac{1+z_r}{\gamma}-1\right)=\frac{z_r-z_b}{2\gamma\beta}.
\label{eq:phizbzr}
\end{equation}
 Finally, by describing the line profile by a Gaussian with an rms width  $ \Sigma(E_0)$ ($E_0$ is the position of the unshifted line centroid), the half-opening angle of the gas flow cone in the jet can be found from the equation 
\begin{equation}
\Theta=\sqrt{\frac{2\ln 2}{3}}~~\frac{2}{\beta\gamma\sin\phi}~~\frac{\Sigma(E_0)}{E_0}.
\label{eq:wtheta}
\end{equation}

As has been noted above, a spectral variability of SS~433 was detected during  \obs\  5513 and 5514 \citep{Marshall2013a}. We check the influence of such a variability on the derived parameters by dividing the exposure of each observation into two parts and repeating the fitting procedure for each part separately. Figure~\ref{fig:beta} shows the results of our measurements for the jet velocity  $\beta$ and opening angle $\Theta$; the circles indicate the result of our analysis for the total exposure, the triangles indicate the results for the two parts of each observation. The blue markers correspond to the parameter measurements from lines in the hard part of the spectrum (6--9 keV) by the method described in Section~\ref{s:hard_part}.

\begin{figure*}
\centering
\includegraphics[width=1\textwidth]{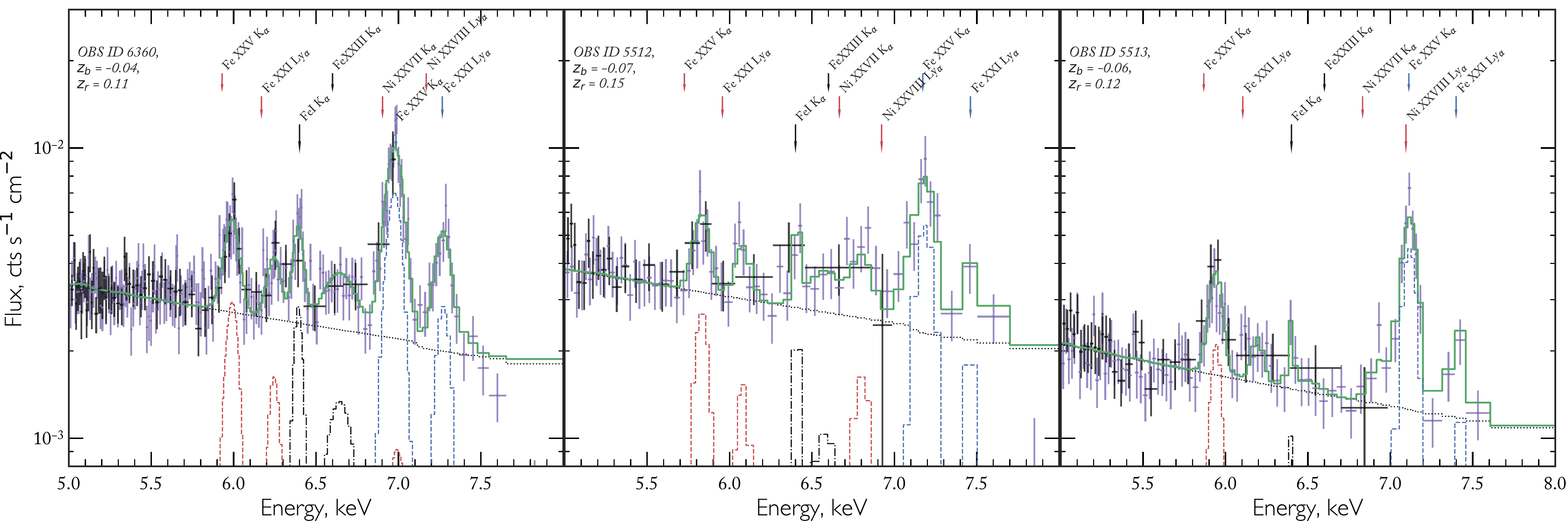}
\caption{\small 
Modeling  \obs\ 6360, 5512, and 5513 in the 5--8 keV energy band using the \texttt{lbjet} jet line emission model (the blue and red lines for the approaching and receding jets, respectively) in combination with the continuum bremsstrahlung model (blue dashes) and two Gaussians (black dashes) with the centroids at 6.4 and 6.6 keV. The figure shows the HEG (blue dots) and MEG (black dots) data binned for better visualization with a detection significance of at least $5 \sigma$. The blue, red, and black arrows at the top indicate the lines of the approaching jet, the receding jet, and the stationary
lines, respectively. }
\label{fig:ljet}
\end{figure*}

\begin{figure*}
\centering
\includegraphics[width=1\textwidth]{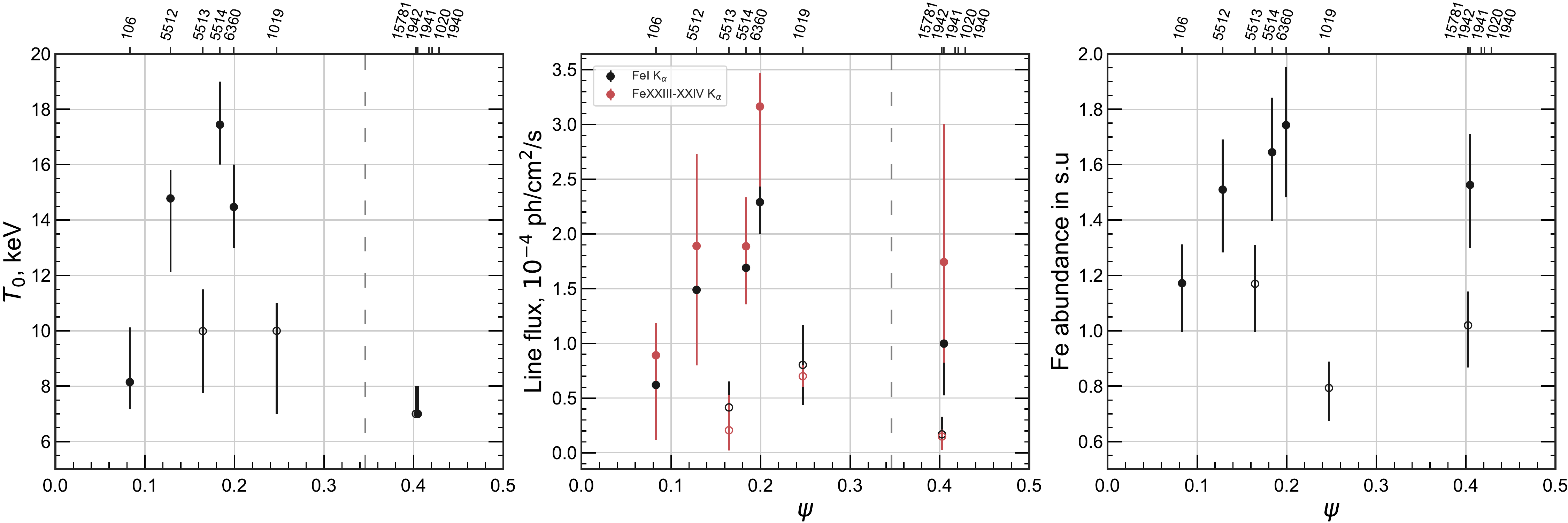}
\caption{\small 
Parameters of the \texttt{lbjet}  model found by analyzing the data in the 5--8 keV energy band versus precession
phase of the system $\psi$; for the observations with phases $\psi>0.5$ the precession phase is mirrored. The vertical dashed lines mark the crossover times. The left panel shows the jet base temperature defining the ratio of the fluxes in the \feK\ and \feLy\ lines. The middle panel presents the fluxes in the fluorescent Fe\,I\,K$_{\alpha}$ line at 6.4 keV (black circles) and in the stationary line at 6.6 keV (red circles), which presumably forms as a result of fluorescence on  Fe\,XXII--XXIII ions in hot wind regions. The right panel shows the iron abundance in solar units \citep{Anders1989} needed to describe the iron lines at fixed parameters of the jet emission model found from the soft part of the spectrum (see Section~\ref{s:soft}).
} 
\label{fig:ljet_params}
\end{figure*}

\section{Emission lines in 5--8 keV energy band}
\label{s:hard_part}
The parameters of the jet emission model found by analyzing the data in the 1--3 keV energy band, namely the total X-ray luminosity of the jets, their base temperature, and the elemental abundances, allow the fluxes in the lines emitted in the spectrum at energies above 3 keV to be predicted. In this section we check whether the inferred jet parameters are consistent with the observed iron and nickel emission lines in the 5--8 keV energy band. For this purpose, we use the approach described in \cite{Medvedev2018} by extracting the continuum emission component from the total spectrum of the  \texttt{bjet} model. Thereafter, the continuum component is subtracted from the total model emission and, thus, the line emission model without a continuum (\texttt{lbjet}) is determined. As a continuum model we use the single-temperature plasma bremsstrahlung spectrum. The fluorescent \feI\  line is described in a standard way, by adding a Gaussian with a fixed centroid at 6.4 keV to the model. Apart from the fluorescent iron line, it is necessary to add one more stationary line with the centroid in the 6.6--6.65 keV energy range, which was also detected from \xmm\ data \citep{Medvedev2018}.

In the first step, the parameters of the \texttt{lbjet}  model were fixed in accordance with those found from the soft energy band; the continuum model parameters were determined by fitting the data. The flux in iron lines in the 5--8 keV energy band estimated in this
way turns out to be a factor of 2 or 3 lower than the observed flux in lines of the hard group of spectra. To get an acceptable quality of the fit, the line normalization in the  \texttt{lbjet} model can be varied by two methods: by varying $L_k \times \tau$, i.e., the total X-ray luminosity of the jets, or by varying the abundances of the corresponding elements. We will dwell on the second method in order to formally determine the difference between the required iron abundances from lines in the soft and hard \chandra\ spectral ranges. For this purpose, we fit the data by releasing the main line parameters in the model (the width and the Doppler shifts, the elemental abundances, and the base temperature), but, at the same time, fixing the X-ray luminosity of the jets found by analyzing the data in the 1--3 keV band. In the case under consideration, the abundance and the temperature are now no longer degenerate, because the ratio of the fluxes in the \feK\ and \feLy\  lines is sensitive to the jet base temperature (see Section~\ref{sec:hard_component}). The best-fit models obtained in this way are shown in Fig.~\ref{fig:ljet}; the derived parameters are shown in Fig.~\ref{fig:ljet_params}.

Our data analysis showed that the jet base temperature needed to describe the lines in the 5--8 keV energy band is slightly higher than that found by
modeling the lines in the soft 1--3 keV energy band (the left panel in Fig.~\ref{fig:ljet_params}). However, to get the observed flux in iron lines, it is necessary to formally increase the iron abundance to  $\approx 1.3$ (with a large scatter), while the iron abundance from the soft part of the spectra is $\approx 0.5$  in solar units. From this it can be concluded that the hotter part of the jets, with a temperature above the base temperature  $T_0$ providing the observed ratio of the line and continuum fluxes in the soft part of the spectrum, must make a significant contribution to the jet emission at energies 3 keV, but, at the same time, be almost completely suppressed in the soft part of the spectrum. The contribution of the emission from such hot and more
compact regions of the jets is confirmed by the observed decrease in the iron line formation temperature at the times of orbital eclipses in the system, while in the soft part of the spectrum no significant changes are observed. As follows from a comparison of the velocities and opening angles from lines in the soft and hard energy bands in Fig.~\ref{fig:beta}, the presumed hot extension of the jets must be completely identical in kinematic and geometric characteristics to the non-eclipsed parts of the jets dominating in the soft part of the spectrum (some deviations were detected only for the times of jet jitter during \obs\  5513 and 5514).

Apart from the jet lines, the fluorescent  \feI\ line is of great interest. There are several versions regarding
the fluorescent emission formation mechanisms in SS~433. \cite{Medvedev2018}  proposed a model in which the fluorescent emission could originate in optically thin regions of the supercritical disk wind, which partially block and scatter the emission from the hottest parts of the jets. Figure~\ref{fig:ljet_params} presents the fluxes in the \feI\ line and the stationary line at $\approx 6.6$  keV. The relatively large measurement error is associated with the model dependence of the contribution of the nickel and iron lines of the receding jet to the observed flux near the lines under consideration. It is clearly seen that the line fluxes change similarly. The line fluxes are also seen to depend on the hardness of the spectra (see Section~\ref{sec:hard_component}); in other words, the flux in the fluorescent line rises as the contribution of the additional component increases. It is clearly seen that both lines, along with the additional component, weaken significantly at the times of orbital eclipses in the system. The fluorescent iron line is described by a narrow Gaussian with a width of 10--20 eV, while the line at 6.6 keV is considerably broader: a line width of
$\sim 60$ eV was obtained for the out-of-eclipse observations; at the times of eclipses the line width decreases to  $\sim 40$ eV. The measured line characteristics, on the whole, are consistent with the model of line formation as a result of fluorescence on iron in the supercritical disk wind. In this case, the line at 6.6 keV must be emitted in the inner, highly ionized parts of the wind, whose velocity must be very high, $\sim 3000$ km s$^{-1}$. Nevertheless, the material in the wind can possibly also expand with higher velocities, as a result of which an observable observed low-contrast broad pedestal is formed in the profiles of optical stationary emission lines  \citep{Medvedev2013}. The position of the line centroid lies in the 6.55--6.64 keV band, which may correspond to the degree of iron ionization  XXII--XXIII.

\section{Hard component model}
\begin{figure}
\centering
\includegraphics[width=1\columnwidth]{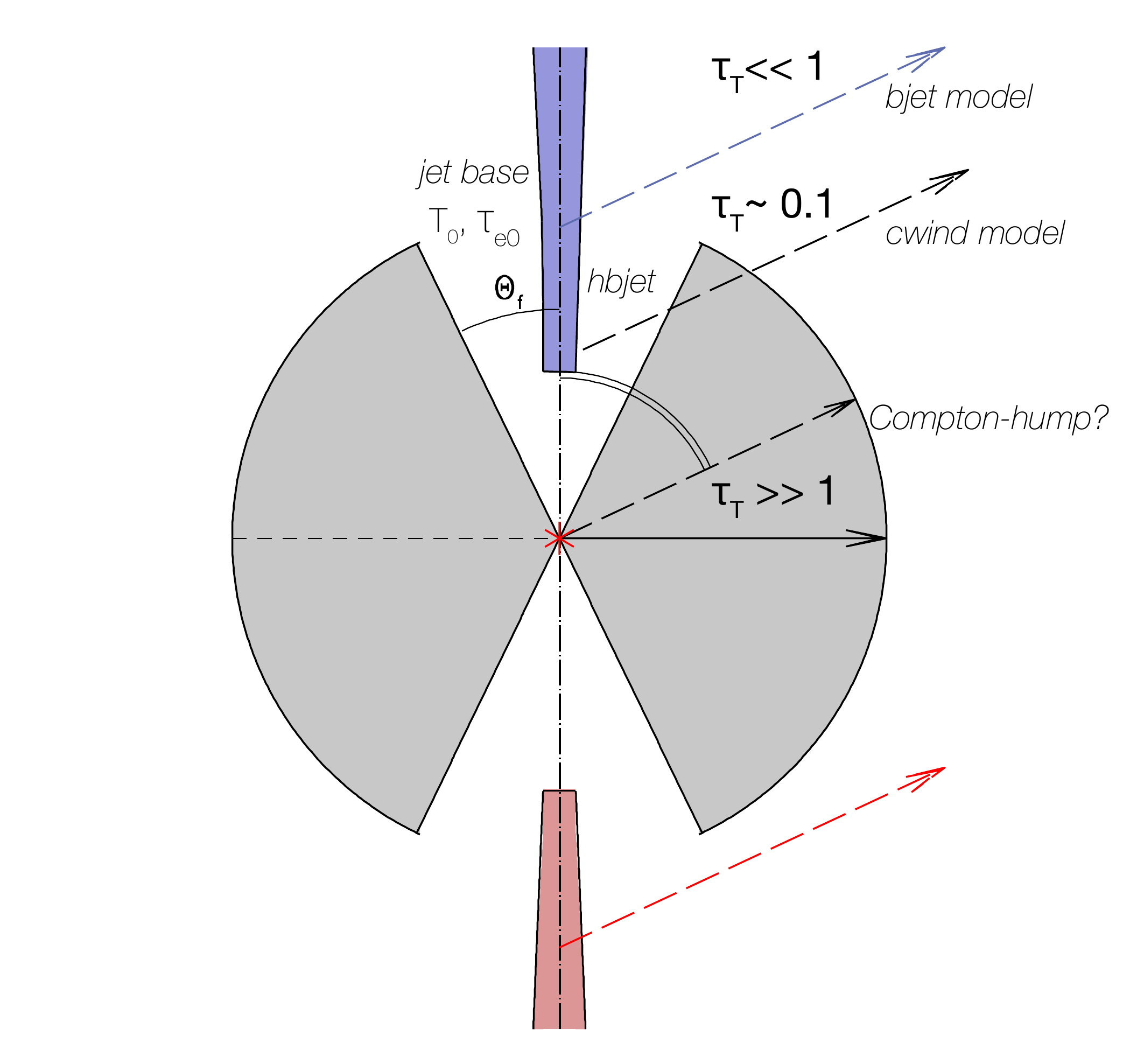}
\caption{\small 
Scheme of the model for the formation of the broadband X-ray spectrum of SS~433 within
the thermal jet emission. The approaching and receding jets are represented by the blue and red regions, respectively. The supercritical accretion disk wind is presented in the form of a homogeneous spherical cloud of neutral gas with a radial Thomson optical depth $\tau_T$ in which
symmetric conical cavities with an opening angle  $\Theta_d = \arccos\mu_d$  were cut out along the axis passing through
the cloud center and forming an angle $ i=\arccos\mu$  with respect to the observer's line of sight.
} 
\label{fig:sketch}
\end{figure}
A high mass transfer rate in the supercritical disk wind ($\sim \dot{M}$) gives rise to dense gas structures around the central engine of SS~433. Obviously, such a wind completely blocks the emission from the jets in the region of their formation, collimation, and acceleration whose size can be compared with the supercritical disk spherization radius  $R_{sph}$ \citep{Poutanen2007}:
\begin{equation}
R_{sph} = \frac{5}{3} \frac{\dot{M}}{\dot{M}_{\rm Edd}}  R_{in} = 1.8\times 10^9\ \text{cm},
\end{equation}
where $R_{in} = 6GM_X/c^2 = 2.7 \times 10^6$ cm is the inner radius of the accretion disk and
$\dot{M}/\dot{M}_{ \rm Edd }\sim\,400$ is the mass transfer rate from the donor star in
units of the Eddington accretion rate  $\dot{M}_{\rm Edd}=3\times 10^{-8} \left(\frac{M_X}{M_{\odot}}\right)\,  M_{\odot}$ yr$^{-1}$, for a mass of the compact object in the system $M_X=3 M_{\odot}$ \citep[see ][]{Fabrika2004}.
At the same time, at distances along the jet axis where the gas temperature approaches the base
temperature, $T_0\approx15$ keV, the wind must have a fairly low density corresponding to an optical depth for electron scattering
$ \lesssim 10^{-2}$, as can be seen from the hydrogen column density estimates (Fig.~\ref{fig:T0}). 
In this case, the geometrical size of this region must be comparable to the effective Roche lobe radius for the donor star in SS~433, i.e.,  $R_{T0} \gtrsim 2\times10^{12}$ \citep{Hillwig2008, Lopez2006, Marshall2013a},  to ensure the absence of changes in the soft part of the spectrum at the times of orbital eclipses. \cite{Medvedev2018}  proposed a scenario in which an excess of hard X-ray emission is formed in the hot extension of the jets whose maximum gas temperature can exceed considerably the visible base temperature $T_0\approx15$ keV.

Such a situation is  possible if there exists a region of the wind around the jets whose density provides an appreciable optical depth for photoabsorption, blocking the jet emission at energies below 3 keV, but, at the same time, optically thin for electron scattering, as a result of which the hard X-ray emission from the hottest parts of the jets passes through and is scattered only partially. In such a situation the fluorescent Fe\,I K$_{\alpha}$  emission line is formed together with the scattered continuum emission whose contribution to the total observed spectrum is expected to be $\tau_T \sim 0.1$. In this paper we continue to develop the idea proposed by \cite{Medvedev2018} by calculating the emission model of the hot extension of baryonic jets (\texttt{hbjet}) intended to become a connecting link between the emission characteristics of the jets, the additional hard component, and the fluorescent emission in the X-ray spectra of SS~433.
 
\subsection{Hot extension of the jets and scattered component}
Within the proposed formalism of the formation of hard X-ray emission the broadband spectrum of SS~433 can be described by a combination of three spectral models (see the scheme in Fig.~\ref{fig:sketch})\footnote{All three models are publicly accessible and can be downloaded from \url{ftp://hea.iki.rssi.ru/medvedev/bjet}}: 
 
\begin{enumerate}
\item the \texttt{bjet} model: the contribution of the emission
from the unabsorbed part of the jets dominating
at energies below 3 keV. A maximum gas temperature
$T_0\approx 15$  keV is reached at a point called the jet base.
The model was proposed by \cite{KMS2016};

\item  the \texttt{hbjet} model: the photoabsorbed emission
from the hot extension of the jets. The contribution
of the component dominates at energies above 3 keV.
The model describes the emission from the part of the
jet with temperatures from $T_0$ to the maximum visible
gas temperature $T_{max}$, after which the jet emission is
completely blocked by the dense wind;

\item the \texttt{cwind} model: the emission component of
the hot extension of the jets scattered in an optically
thin wind with the fluorescence of neutral atoms.
The component reproduces the observed flux in fluorescent
lines, but the integrated contribution of the
component is relatively small, $\sim \tau_T \sim 0.1$. The model
was proposed by \cite{Medvedev2018}.

\end{enumerate}

The \texttt{hbjet} spectral model was computed by analogy with the  \texttt{bjet} model   \citep{KMS2016}. The boundary (initial) conditions for solving the thermal balance equation (see Eq.~\ref{eq:cooling}) are specified as before at a point along the jet axis called the base ($r_0$), where the jet is directly visible to the observer. Thereafter, the thermal balance equation is solved in the ``backward'' direction, in the direction of increasing gas temperature. The computation is terminated at $r_{min}$, when the gas temperature reaches $T_{max}$. We assume that the emission from a portion of the jet with $r<r_{min}$ is completely blocked by the accretion disk wind. The model was computed for a similar set of parameters as that for the \texttt{bjet} model, but it has an additional parameter for the maximum visible gas temperature $T_{max}$.

The \texttt{cwind}  spectral model is described in \cite{Medvedev2018}.
In this model the emission source is placed at the center of a homogeneous spherical cloud of neutral gas with a radial Thomson optical depth $\tau_T$ and a heavy-element abundance $Z$, in which symmetric conical cavities with an opening angle $\Theta_d = \arccos\mu_d$  were cut out along the axis passing through the cloud center and forming an angle $ i=\arccos\mu $ with the observer's line of sight (see Fig..~\ref{fig:sketch}). The point source is assumed to correspond to the part of the jet with the highest temperature $T_{max}$ whose spectrum is specified by the single-temperature bremsstrahlung model.

\subsection{Fitting results}
The parameters of the  \texttt{bjet} model are fixed in accordance with those found by analyzing the soft part of the spectrum (see Section~\ref{s:fitting}). Since obtaining accurate parameters of the model for the hard component is beyond the scope of this paper, for a qualitative description of the spectrum we use the  \texttt{hbjet} model only for the approaching jet, which makes a major contribution to the observed spectrum. This approximation seems reasonable, because for the observations at precession phases $\psi < 0.2$  the suppression factor of the opposite jet $constant$  found from our fitting points to a very significant additional absorption probably attributable to the gas in the system's equatorial plane. The  \texttt{hbjet} emission photoabsorption, as before, is specified by the  \texttt{phabs} model. All of the model parameters, except for the maximum gas
temperature $T_{max}$, are joined at $r_0$ corresponding to the position of the jet base in the  \texttt{bjet} model. For
the \texttt{cwind} model of the scattered emission component the source's bremsstrahlung temperature is related to $T_{max}$  and the optical depth is related to the parameter $N_H$ of the multiplicative  \texttt{phabs} model for the hot part of the jet  \texttt{hbjet}. The derived model can be written as
\begin{align*}
\small
& \textsf{Model} = phabs* (gsmooth * (zashift_{b}*bjet_{b} +\\
& + constant*zashift_{r}*bjet_r )  + \\ 
&  + phabs*gsmooth *zashift_{b}*hbjet_{b}+\\
& + cwind).
\label{eq:model}
\end{align*}
The derived best-fit models for \obs\ 6360 and 5513 are shown in Fig.~\ref{fig:hjet}. We see that
there is a deficit of the flux in iron lines in the 5--8 keV energy band; we think that this may stem from
the fact that the model used is obviously simplified. In particular, we disregard the change in the density of wind material along the line of sight for different
points along the jet axis by fixing the parameter $N_H$ for the entire emergent radiation of the  \texttt{hbjet} model.
Besides, a neutral and homogeneous gas in a wind with a simple spherically symmetric geometry is considered in the \texttt{cwind}  model, which, obviously, is a rough approximation. The central part of the wind can have a high degree of gas ionization, as suggested by
the observed stationary line at 6.6 keV described in Section~\ref{s:fitting}. The wind is also expected to have a significantly inhomogeneous structure \citep[see, e.g., ][]{Ohsuga2011}. Nevertheless, we see that the proposed model can qualitatively reproduce the hard continuum at energies $>3$ keV and compensate almost completely for the flux excess in iron lines.

\section{Conclusions} 
\label{s:discussion}
The unknown spectral shape of the emission from the additional component introduces great uncertainties in describing the broadband \chandra\  spectrum by the jet emission model. We made an attempt to separate the X-ray emission produced in the jets from the emission of the additional component through a systematic data analysis. For this purpose, we used data in the soft 1--3 keV X-ray energy band, where the bulk of the X-ray lines are emitted and the contribution of the additional component is expected to be comparatively small. In this part of the spectrum the energy resolution and sensitivity of the HETGS diffraction gratings reach the highest values, which allows the line intensities and positions for the approaching and receding jets to be measured separately. Such measurements give valuable quantitative information about the temperature and chemical composition of the gas in the jets. However, the elemental abundances being determined turn to to be directly related to the presumed temperature of the visible jet base, which leads to degeneracy of the model parameters when directly fitting the data. A constraint on the jet base temperature can be obtained by analyzing the relative intensities of the lines originating in jet regions with different temperatures providing the maximum plasma emissivity in a given line. In practice, unfortunately, this method is unsuitable when analyzing the lines in the soft part of the spectrum, because the emission measure distribution in the cold ``tail'' of the X-ray jets is insensitive to the jet base temperature. On the other hand, such a situation allows the relative abundances of elements whose lines are emitted in the soft X-ray energy band to be accurately measured. Our data analysis showed that the relative abundances of all elements, except for nickel, are close to their solar values. This allowed us to determine the jet base temperature, $T_0\approx15$, at which the continuum level in the soft part provides the observed line equivalent widths under the condition of a nearly solar chemical composition of the jets.

The model parameters found from the soft part of the spectrum allow the fluxes in jet emission lines to be predicted in the entire \chandra\ energy range. The observed fluxes and the required jet base temperature for the iron lines in the 5--8 keV energy band turned out to be noticeably higher than those predicted in the model from the lines in the soft part of the spectrum. In contrast to the lines in the soft part of the spectrum, the observed fluxes in the iron lines point to a decrease in the jet base temperature at the times of orbital eclipses in the system.

As an explanation of the additional component we develop the formalism proposed in \cite{Medvedev2018}. According to the scenario proposed in this paper, an excess of hard X-ray emission is formed in the hot extension of the jets in which the maximum gas temperature can significantly exceed the visible base temperature $T_0\approx10$ . Such a situation is possible if there exists a region of the wind around the jets whose density provides an appreciable optical depth for photoabsorption, blocking the jet emission at energies below 3 keV, but, at the same time, optically thin for electron scattering, as a result of which the hard X-ray emission from the hottest parts of the jets passes through and is scattered only partially. In this paper we computed the emission model of the hot extension of baryonic jets (\texttt{hbjet} model) intended to become a connecting link between the emission characteristics of the jets, the additional hard component, and the fluorescent emission in the X-ray spectra of SS 433. The parameters of the models for the absorbed and unabsorbed parts of the jets are linked by the boundary conditions at the jet base, where the jet becomes directly visible to an observer without absorption in the wind. A combination of the emission model for the unabsorbed X-ray jets with the partially scattered and photoabsorbed emission from the hot extension of the jets gives a complete self-consistent picture of the formation of the broadband X-ray spectrum for SS 433 within the thermal emission of the baryonic jets in the system. The required equivalent hydrogen column density for the photoabsorbed part of the jets turns out to be within the reasonable range $N_H = 15$--$20\times 10^{22}$ atoms cm$^{-2}$ with a maximum
gas temperature up to 40 keV.

\section*{Acknowledgements}
The research was supported by the Russian Science Foundation (project no. 14-12-01315).
We thank E.~M.~Churazov for the provided \texttt{cwind} spectral model of scattering in a cold wind.

\bibliographystyle{mnras}
\bibliography{biometrics_en}

\clearpage
\thispagestyle{empty}
\onecolumn
\begin{landscape}
\begin{figure*}
\centering
\includegraphics[width=1.35\textwidth]{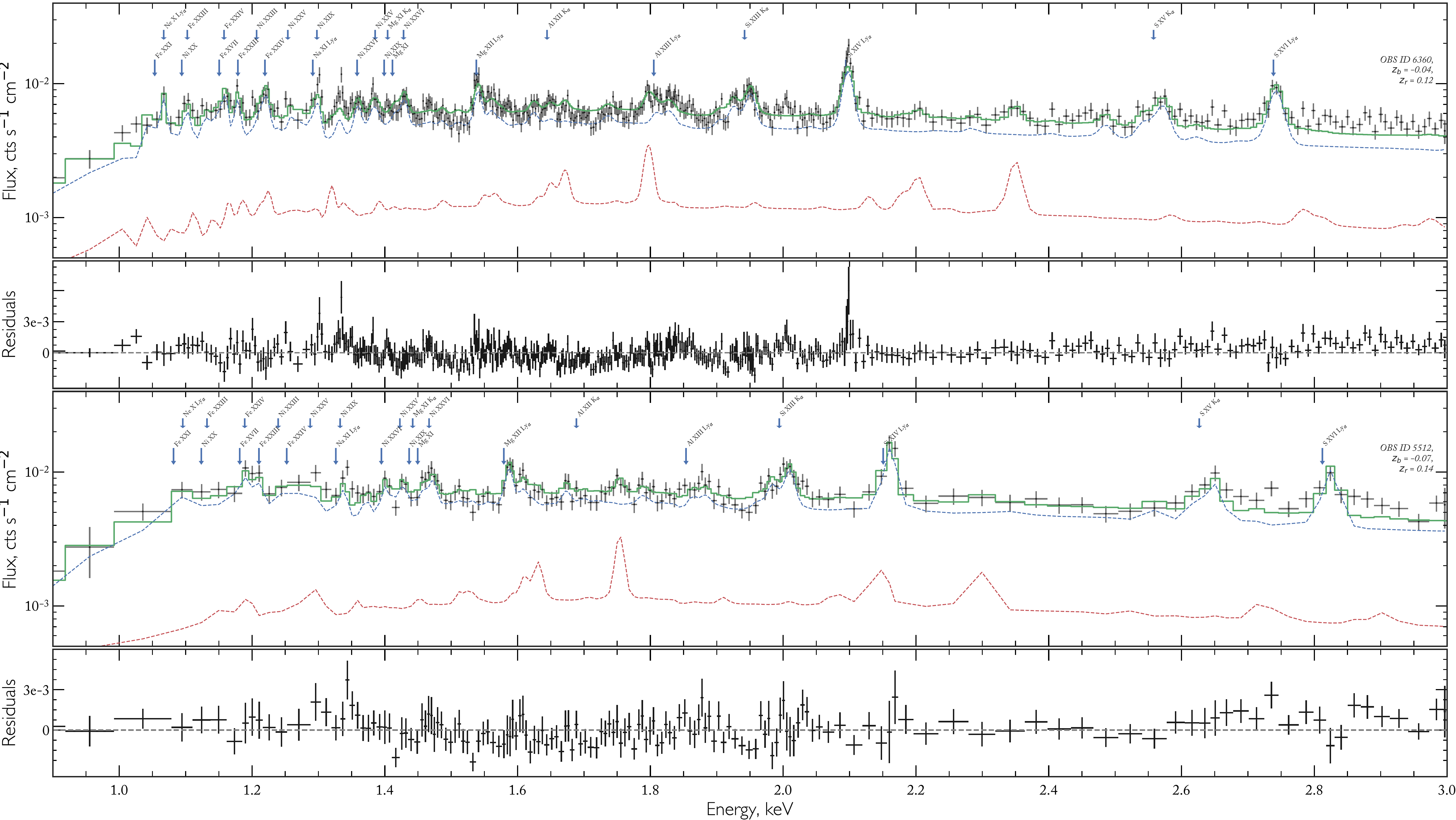}
\caption{\small 
Results of fitting the HEG and MEG data for  \obs\  6360 (two upper panels) and \obs\  5512 (two lower panels) by the thermal jet emission model in the 1--3 keV energy band. The figure shows only the MEG data, the spectral channels were binned for better visualization, each bin has a detection significance of at least $7 \sigma$. The combined model is indicated by the green solid curve; the emission from the approaching and receding jets is indicated by the blue and red dashed lines, respectively. The derived Doppler shifts of the jets are: \obs\  6360: 
$z_b = -4.40\pm0.02 \times 10^{-2} $, $z_r = 11.57\pm0.09 \times 10^{-2}$ and \obs\ 5512: $z_b = -7.21\pm0.03 \times 10^{-2} $, $z_r = 14.32\pm0.14 \times 10^{-2}$. The  K$_{\alpha}$ and Ly$_{\alpha}$ emission lines of hydrogen- and helium-like Ne, Na, Mg, Al, Si, and S ions as well as the most intense iron and nickel lines in the 1--1.5 keV band of the approaching jet are indicated at the top.
} 
\label{fig:fit6360}
\end{figure*}
\end{landscape}
\clearpage
\thispagestyle{empty}
\onecolumn
\begin{landscape}
\begin{figure*}
\centering
\includegraphics[width=1.35\textwidth]{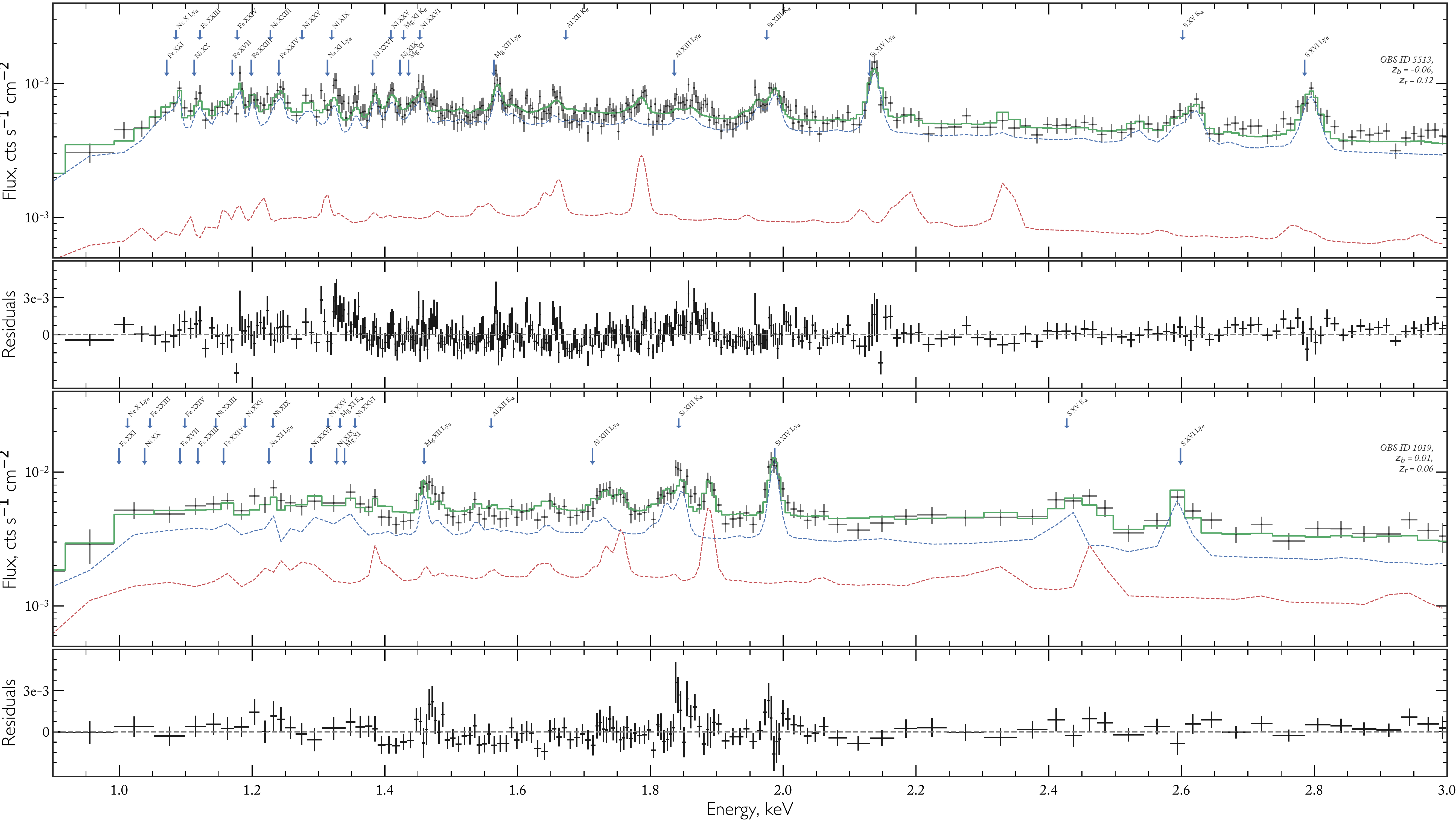}
\caption{\small 
Same as Fig.~\ref{fig:fit6360} for  \obs\ 5513 and \obs\ 1019. The observations correspond to the times of deep orbital eclipses (orbital phase $\phi \approx 0$), but different precession phases of the system: $\psi=0.16$ for \obs\ 5513 is a phase at which the accretion disk is open almost completely to the observer and $\psi=0.26$ for \obs\ 1019 is a phase close to the crossover, when the observer views the accretion disk edge-on. The derived Doppler shifts of the jets are: \obs\ 5513: $z_b = -6.20\pm0.05 \times 10^{-2} $, $z_r = 13.39\pm0.17 \times 10^{-2}$  and \obs\ 1019:  $z_b = 0.93\pm0.05 \times 10^{-2} $, $z_r = 6.2\pm0.07 \times 10^{-2}$.
} 
\label{fig:fit5513}
\end{figure*}
\end{landscape}

\clearpage
\thispagestyle{empty}
\onecolumn
\begin{landscape}
\begin{figure*}
\centering
\includegraphics[width=1.35\textwidth]{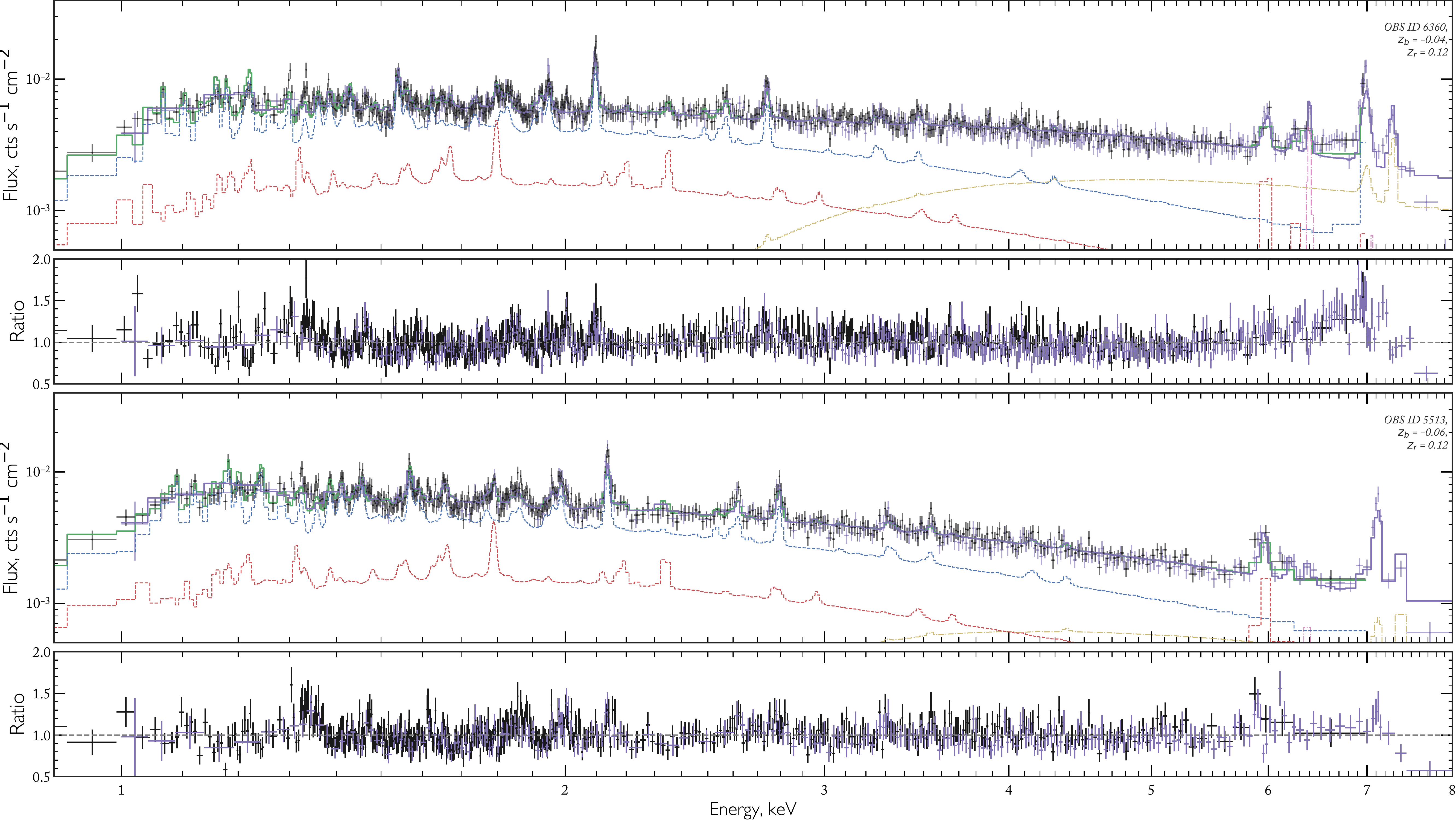}
\caption{\small 
Modeling the additional hard component in the Chandra spectra of SS~433 (\obs\ 6360 and 5513). The figure shows the HEG (violet dots) and
MEG (black dots) data binned for better visualization with a detection significance of at least $7 \sigma$. The combined model is indicated by the solid green and violet curves (for the MEG and HEG, respectively); the components of the approaching and receding jets (\texttt{bjet} model) are represented by the blue and red dashed lines, respectively. The contribution of the emission from the absorbed hot extension of the approaching jet (\texttt{hbjet} model) is indicated by the yellow dashed line. The scattered emission component from the hottest parts of the jet (\texttt{cwind} model) is indicated by the pink dashed line (fluorescent iron line).} 
\label{fig:hjet}
\end{figure*}
\end{landscape}
\twocolumn

\end{document}